\documentclass[11pt,a4paper]{article}
\pdfoutput=1
\usepackage{jheppub}

\usepackage{epsf,epsfig,amsmath,amssymb,dsfont}
\usepackage{amsthm,mathrsfs} 
\usepackage{graphicx} 
\usepackage{multirow}
\usepackage{float}
\usepackage{colortbl}
\usepackage[table]{xcolor}
\usepackage{varioref}
\usepackage{slashed}
\usepackage{float}

\newcommand{\be}{\begin{equation}}
\newcommand{\ee}{\end{equation}}
\newcommand{\baln}{\begin{align}}
\newcommand{\ealn}{\end{align}}
\newcommand{\ben}{\begin{equation*}}
\newcommand{\een}{\end{equation*}}

\long\def\symbolfootnote[#1]#2{\begingroup%
\def\thefootnote{\fnsymbol{footnote}}\footnote[#1]{#2}\endgroup}

\title{Lorentz violation naturalness revisited}

\author[a,b]{Alessio Belenchia,} 
\author[a,b]{Andrea Gambassi}
\author[a,b]{and Stefano Liberati}

\affiliation[a]{SISSA - International School for Advanced Studies, via Bonomea 265, 34136 Trieste, Italy.}
\affiliation[b]{INFN, Sezione di Trieste, Trieste, Italy.}

\emailAdd{abelen@sissa.it}
\emailAdd{gambassi@sissa.it}
\emailAdd{liberati@sissa.it}

\abstract{We revisit here the naturalness problem of Lorentz invariance violations on a simple toy model of a scalar field coupled to a fermion field via a Yukawa interaction. We first review some well-known results concerning the low-energy percolation of Lorentz violation from high energies, presenting some details of the analysis not explicitly discussed in the literature and discussing some previously unnoticed subtleties. We then show how a separation between the scale of validity of the effective field theory and that one of Lorentz invariance violations can hinder this low-energy percolation. While such protection mechanism was previously considered in the literature, we provide here a simple illustration of how it works and of its general features. Finally, we consider a case in which dissipation is present, showing that the dissipative behaviour does not percolate generically to lower mass dimension operators albeit dispersion does. Moreover, we show that a scale separation can protect from unsuppressed low-energy percolation also in this case.
}
\begin{document}
\maketitle
\flushbottom


\section{Introduction}\label{introduction}
Symmetries play a fundamental role in theoretical physics. In particular, spacetime symmetries are at the basis of quantum field theory (QFT) and general relativity (GR), the two pillars of modern physics. It is then natural to investigate and question the very nature of these symmetries, e.g. whether they are exact or accidental, i.e. emerging in the low-energy world that we can access with present experiments. (Local) Lorentz invariance (LI) is one of the best tested symmetries of Nature even if the Lorentz group, being non-compact, makes its probing process an endless task. As far as we can say, Lorentz symmetry presently appears to be an exact symmetry since tests of Lorentz invariance violations (LIV) have provided stringent bounds on possible violations~\cite{Liberati:2013xla,Mattingly:2005re,Amelino-Camelia2013}. So, in principle, one could wonder about the point in questioning this symmetry. The answer, as it is well known, comes from the fact that various quantum gravity (QG) proposals seem to entail violations (or modifications) of Lorentz symmetry at the fundamental level~\cite{Douglas:2001ba,Gambini:1998it,Urrutia:2005ty,Mavromatos:2007xe,Horava:2009uw,Jacobson:2008aj,AmelinoCamelia:2000ge,AmelinoCamelia:2011bm} \footnote{See, however, ref.~\cite{Rovelli:2010ed} for what concerns loop quantum gravity.}. It is then interesting to test if such violations can be detected or alternatively used to rule out some QG scenario. In fact, any viable theory of quantum gravity/spacetime needs to treat carefully LI in order to recover it in the low-energy limit. This is far from being a trivial task. Among the various QG theories there are some that are not affected by this problem since they assume LI from the outset, as in the case of causal-set theory \cite{Bombelli:1987aa,Bombelli:2006nm}, while other proposals modify the action of the Lorentz group by making it non-linear \cite{Amelino-Camelia2013}. It is worth mentioning that in the latter approach, known as doubly--special relativity (DSR)~\cite{AmelinoCamelia:2000mn}, modified dispersions relations such as the ones considered in the following, are expression of an extended symmetry group. Accordingly, naturalness arguments typical of LIV effective field theory do not straightforwardly apply.

For what concerns other approaches to QG the quest for how to recover LI at low energies should be of primary importance and it is, in many cases, an open issue.  
Unfortunately, even if LIV is conjectured only in the far ultra-violet (UV)  completion of QFT (i.e. at Planckian scales), still it would lead to large low-energy effects.
In a seminal work \cite{Collins:2004bp}, Collins \textit{et al.} showed that in a generic QFT, seen as an effective field theory (EFT), Lorentz violations in the UV can percolate in the infra-red (IR) without being suppressed and leading to unacceptably large effects. Here the term ``percolation'' refers to the fact that in an EFT setting even if one starts by adding only LIV operators of mass dimension larger than four, radiative corrections will generate mass dimension four (and, generically, mass dimension three) operators. Then the percolation is said to be unsuppressed if there is no small amplitude suppressing the effects of these operators (in addition to usual coupling constants). This is a peculiarity of LIV. Indeed, in the case of LI theories, the physics at high energies affects the IR physics only via renormalization of the bare couplings of the theory. Instead, in the presence of LIV in the UV these effects can percolate unsuppressed in the IR, through radiative corrections. This can be easily understood from the EFT point of view. Indeed, in EFT every operator respecting the fundamental symmetries of the theory can be present. Accordingly, even if a certain operator is not present in the UV, radiative corrections can give rise to mass dimension four (and in general three) Lorentz invariance violating operators (see also refs.~\cite{Iengo:2009ix,Gambini2011,Polchinski2012}) once higher-order LIV operators are allowed in the theory.

The conclusion of ref.~\cite{Collins:2004bp} was that if LIV is admitted, then it is necessary to accept an unnaturally and extreme fine-tuning of the theory, in order to respect the stringent experimental constraints on LIV, otherwise the theory would be immediately ruled out by observations. From this point of view, LIV can be considered a new fine-tuning problem to be added to the various ones existing in particle physics and cosmology.

Clearly, various solutions to this fine-tuning problem have been proposed in the literture (see ref.~\cite{Liberati:2013xla} for an extended discussion and ref.~\cite{Afshordi:2015smm} for recent developments). One route relies on custodial symmetries suppressing the percolation in the IR. This is the case of supersymmetry and in general of other models including a spontaneous symmetry breaking~\cite{GrootNibbelink:2004za,Bolokhov:2005cj, Sindoni:2007rh} which can serve as a paradigmatic example of the possibility of new (a priori Lorentz invariant) physics between the electroweak scale and an eventual Lorentz breaking at the Planck scale. This is the case we shall be mostly concerned within this work.

Another possible solution to the LIV fine-tuning problem is to restrict the violation to the gravity sector of the theory, leaving the matter sector LI, as in the case of the \textit{gravitational confinement} proposed by Pospelov and Shang~\cite{Pospelov:2010mp}. In particular, a separation between the Planck and the LIV scale in the gravity sector (with the latter taken to be smaller than the former) was used to show how percolation can be tamed. This proposal relies on the Planck mass suppressed vertices that appear in the matter sector due to the coupling with gravity and on the fact that LIV in the gravitational sector is not so stringently tested~\cite{Pospelov:2010mp}. 

Finally, a third possible way of achieving an infrared protection from high-energy LIV is the one envisaged by Nielsen and collaborators in the seventies, which makes use of renormalization-group techniques~\cite{nielsen1978beta,Nielsen:1982kx}. While the standard logarithmic running towards a LI theory in the infrared is generically not fast enough to be compatible with current observations, it was nonetheless noticed that a strong coupling close to the Planck scale can sufficiently enhance the running such that almost exact LI is rapidly achieved. This is the basic idea behind the proposal of ref.~\cite{Bednik:2013nxa}.

Before proceeding, let us mention as a cautionary note, that the low-energy effective field theory paradigm, as well as the related naturalness arguments, are not always capable of capturing the correct physics. As an example, consider the effective field theory prediction for the magnitude of the cosmological constant, which is off by more than 120 orders of magnitude compared with observations. Without a direct measurement, one would have expected a naturalness problem for the cosmological constant. The observational evidence that the latter has such a small value seems to suggest, instead, a breakdown of EFT or the presence of yet to be understood symmetries at intermediate energies between the TeV and the Planck scale. In the following we will focus only on the EFT description but keeping in mind that a breakdown of an EFT-based intuition might apply also in the case of LIV naturalness. 

In this work we first review the argument of refs.~\cite{Collins:2004bp,Collins2006}, presenting some calculations which are not available in the literature and showing how, in some special cases, a cancellation of the percolation can be achieved. Then, we attack the problem from a different perspective, illustrating in a simple way how the first of the above proposal works. We shall indeed show that, in the case in which there is a large separation between the EFT validity scale $\Lambda$ and the LIV scale $M$, the percolation can be suppressed and discuss how this suppression generically scales with energy. In particular, we find that for dispersion relations with leading LIV terms of order $(E/M)^{2n}$ the percolation scales as $\Delta c\propto (\Lambda/M)^{2n}$. This implies that if any such scenario could be successfully applied to the SM, values of $\Lambda$ below $10^{10}$ GeV would be sufficient to reconcile the most interesting (CPT invariant) case $n=1$ with current observational constraints (see further discussion in sections \ref{sec4} and \ref{sec6}).
Finally, we also consider a dissipative case, in the spirit of ref.~\cite{Parentani:2007uq}, and show that while the dissipative behaviour (i.e.~the presence of imaginary contributions) does not percolate, a dispersive one does. We also demonstrate that a scale separation can hinder such percolation with basically the same behaviour as the one we find in the dispersive case.

The paper is organised as follows: After a brief introduction, in section~\ref{sec2} we review the work by Collins \textit{et al.}~\cite{Collins:2004bp}, presenting some details of the calculation. In section~\ref{section3} we calculate the fermion self energy at one loop and use it in order to gain information on the percolation of LIV on the fermion. In section~\ref{sec4} we introduce a second scale in the problem in the form of as a sharp or smooth LI cutoff and show that the situation changes significantly. In section~\ref{sec5} we consider what happens if the LIV is associated with dissipative phenomena. Finally we conclude with a discussion in section~\ref{sec6}.
 
\section{LIV percolation: previous results\label{sec2}}

In this section we briefly review, for the readers' convenience, the results of ref.~\cite{Collins:2004bp}. This work considers a model of a scalar $\phi$ and a fermion $\psi$ coupled via a Yukawa interaction
\begin{equation}\label{lag}
\mathcal{L}=\frac{1}{2}(\partial\phi)^2 -\frac{m_{\phi}^{2}}{2}\phi^2 +\bar{\psi}(i\gamma^{\mu}\partial_{\mu}-m_{\psi})\psi+g\phi\bar{\psi}\psi,
\end{equation}
where $g$ is the dimensionless coupling constant\footnote{Hereafter we consider natural units, with $c=1$.}. Beyond tree-level, the theory is made finite by a cutoff on spatial momenta (in a given preferred frame), implemented as a modification of the free propagators. As a first important point, we emphasise here that the scale entering as a (LIV) cutoff for the UV behaviour of the theory is the LIV scale itself, i.e. the same scale as the one appearing in the modified dispersion relations (MDR). Moreover, this scale can be identified with the Planck scale. Once such a cutoff is introduced, one assumes for the scalar and fermion propagators in momentum space (see refs.~\cite{Collins:2004bp,Collins2006} for details) 
\begin{align}
& \frac{i}{\slashed{p}-m_{\psi}+i\epsilon}\rightarrow\frac{i f\left(|\textbf{p}|/\Lambda\right)}{\slashed{p}-m_{\psi}+\Delta\left(|\textbf{p}|,\Lambda\right)+i\epsilon}, \label{pro1}\\
& \frac{i}{p^{2}-m_{\phi}^{2}+i\epsilon}\rightarrow\frac{i \tilde{f}\left(|\textbf{p}|/\Lambda\right)}{p^{2}-m_{\phi}^{2}+\tilde{\Delta}\left(|\textbf{p}|,\Lambda\right)+i\epsilon}, \label{pro2}
\end{align}
where $|\textbf{p}|$ is the modulus of the 3-momentum and the cutoff functions $f\left(|\textbf{p}|/\Lambda\right)$ and $\tilde{f}\left(|\textbf{p}|/\Lambda\right)$ are such that they approach 1 as $|\textbf{p}|/\Lambda \ll 1$, in order to reproduce low-energy physics, while they vanish sufficiently rapidly as $|\textbf{p}|/\Lambda\gg 1$, in order to render the theory UV finite. The functions $\tilde{\Delta}$ and $\Delta$ appearing in the denominators, instead, come from actual proposals for MDR that usually appear in the quantum gravity (phenomenology) literature~\cite{Liberati:2013xla, Amelino-Camelia2013, Mattingly:2005re}. These functions are such that they vanish for $|\textbf{p}|/\Lambda\ll 1$, again to recover the low-energy physics. Actually, in ref.~\cite{Collins:2004bp}, $\Delta$ and $\tilde{\Delta}$ are not introduced since it is argued that they will not affect the argument. This is due to the fact that the LIV effects are seen as producing a natural cutoff at large (spatial) momenta which is already produced by the cutoff  functions $f$ and $\tilde{f}$. However, note that neglecting $\Delta$ and $\tilde{\Delta}$, i.e. the MDR, is possible only at the price of identifying the EFT and the MDR scales. This is tantamount to assuming no new physics between the Standard Model (SM) scale and the Planck one. 

It has been objected that introducing a new scale between the low-energy scale and the Planck scale would be intrinsically against the philosophy of the quantum gravity phenomenology models entailing Lorentz breaking in the UV, which hinge on the persistence at low energies of these Planck scale effects~\cite{Perez:2003un}. However, we do not see a problem in conjecturing such a hierarchy of scales (we rather found difficult to conceive that no new physics will be present from the TeV to the Planck scale): the effective field theory framework is perfectly capable to account for these scenarios. Of course, one can conceive that within this scenario additional operators which are suppressed in the EFT scale might be present at the tree level but we notice that most of the phenomenological constraints on Lorentz violations are based on anomalous mechanisms (new threshold reactions or upper thresholds) which cannot be generated otherwise (i.e.~they are intrinsically preferred frame effects). For these reasons, later on in this work we will drop such an identification by requiring the LIV scale which enters the MDR to be different from the cutoff scale of the theory that we will introduce in a Lorentz-invariant way. Let us stress that this does not imply any notion of intermediate Lorentz invariance between the Planck scale and currently tested energies. Indeed, in the scenario considered here all the physics below the Planck scale is Lorentz breaking.  What we are envisaging here is that there could be some exact symmetry of nature, such as~SUSY, which could be broken below some energy scale $\Lambda\ll M$ and that the new physics associated with this symmetry is not per se inducing Lorentz breaking operators suppressed by a scale other than $M$ (so for $M\to \infty$ all the physics should be LI).

The unsuppressed percolation of LIV from the UV to the IR was considered in ref.~\cite{Collins:2004bp} on the scalar field by computing its one-loop self energy. As anticipated, we briefly review below the full computation. In the presence of LIV, the inverse propagator $\Pi$ of the scalar field including one-loop corrections can be written as 
\begin{equation}\label{sel}
\Pi(p)=A+B p^2 +\xi p^{\mu}p^{\nu}W_{\mu}W_{\nu}+\Pi^{(LI)}(p^2)+\mathcal{O}(p^4/\Lambda^2),
\end{equation}
where $W^{\mu}$ is a unit timelike background vector field permitting to write a LIV expression in a covariant form\footnote{Recall that when speaking of Lorentz invariance in field theory we always refer to active Lorentz invariance, see the discussion in ref.~\cite{Mattingly:2005re}.}. The third term on the r.h.s. of eq.~\eqref{sel} contains the unsuppressed LIV, i.e. it results in a different limit velocity of the scalar field\footnote{The explicit value of the parameter $\xi$ resulting from LIV will be dependent on the coupling constant of the theory, as it is due to radiative corrections. Then, considering more rich theories one can see that the different particles will have different limit speeds in such a way that a simple rescaling of the limit velocity cannot eliminate this effect.} (with respect to $c$) while $\Pi^{(LI)}$ is a Lorentz invariant term (see ref.~\cite{Collins2006}). Our aim here is to compute $\xi$.  

In what follows we will use the propagators in eqs.~\eqref{pro1} and~\eqref{pro2}, this means that we set the calculation in the preferred reference frame characterised by the 4-velocity $W^{\mu}$. In this case, the coefficient $\xi$ of interest is obtained as
\begin{equation}\label{cscalar}
\xi =\left[\frac{\partial^{2}\Pi}{\partial (p^{0})^{2}}+\frac{\partial^{2}\Pi}{\partial (p^{1})^{2}}\right]_{p=0}.
\end{equation} 
In section \ref{21} we briefly consider for pedagogical purpose the LI case, while in section \ref{dd} we consider the LIV case originally studied in ref.~\cite{Collins:2004bp}.
\subsection{LI case}\label{21}
When there is no LIV, $\xi$ is expected to vanish. In fact, this can be seen directly from eq.~\eqref{cscalar}. Computing the diagram in figure \ref{loop1}, we find  

\begin{align}
\Pi(p^2)&=-i g^2 \int\frac{d^{4}k}{(2\pi)^{4}}\frac{tr[(\slashed{k}-\slashed{p}+m_{\psi})(\slashed{k}+m_{\psi})]}{[(k-p)^2 -m_{\psi}^{2} ][k^2 -m_{\psi}^{2} ]}\\ \nonumber
&= -i g^2 \int\frac{d^{4}k}{(2\pi)^{4}}\frac{4 (k^2 -p\cdot k)+4m_{\psi}^{2} }{[(k-p)^2 -m_{\psi}^{2}][k^2 -m_{\psi}^{2} ]},
\end{align}
and so 
\begin{align}\label{indiv}
\xi &=\left[\frac{\partial^{2}\Pi}{\partial (p^{0})^{2}}+\frac{\partial^{2}\Pi}{\partial (p^{1})^{2}}\right]_{p=0}\\ \nonumber
&=\frac{-ig^2}{\pi^4}\int d^{4}k\frac{[(k^0)^2+(k^1)^2](k^2+3m^2)}{(k^2-m^2)^4},
\end{align}
where hereafter we set $m\equiv m_{\psi}$ to simplify the notation.
Although the integral in eq.~\eqref{indiv} is formally logarithmically divergent (by power counting), the fact that it actually vanishes can be understood from a symmetry argument~\cite{Collins:2004bp}. Indeed rotating in the Euclidean space and using four-dimensional spherical symmetry it is straightforward to see that the angular part of the integral implies $\xi=0$.


\subsection{LIV case}\label{dd}
We present now some details of the computation in the presence of LIV. We will relegate some technicalities to appendices \ref{appA} and \ref{appB} while showing here the main steps in a self-contained way. We will work from now on with mostly minus signature.  

\begin{figure}[tbp]
\centering
\includegraphics[trim={0 9cm 0 9cm},clip, scale=0.4]{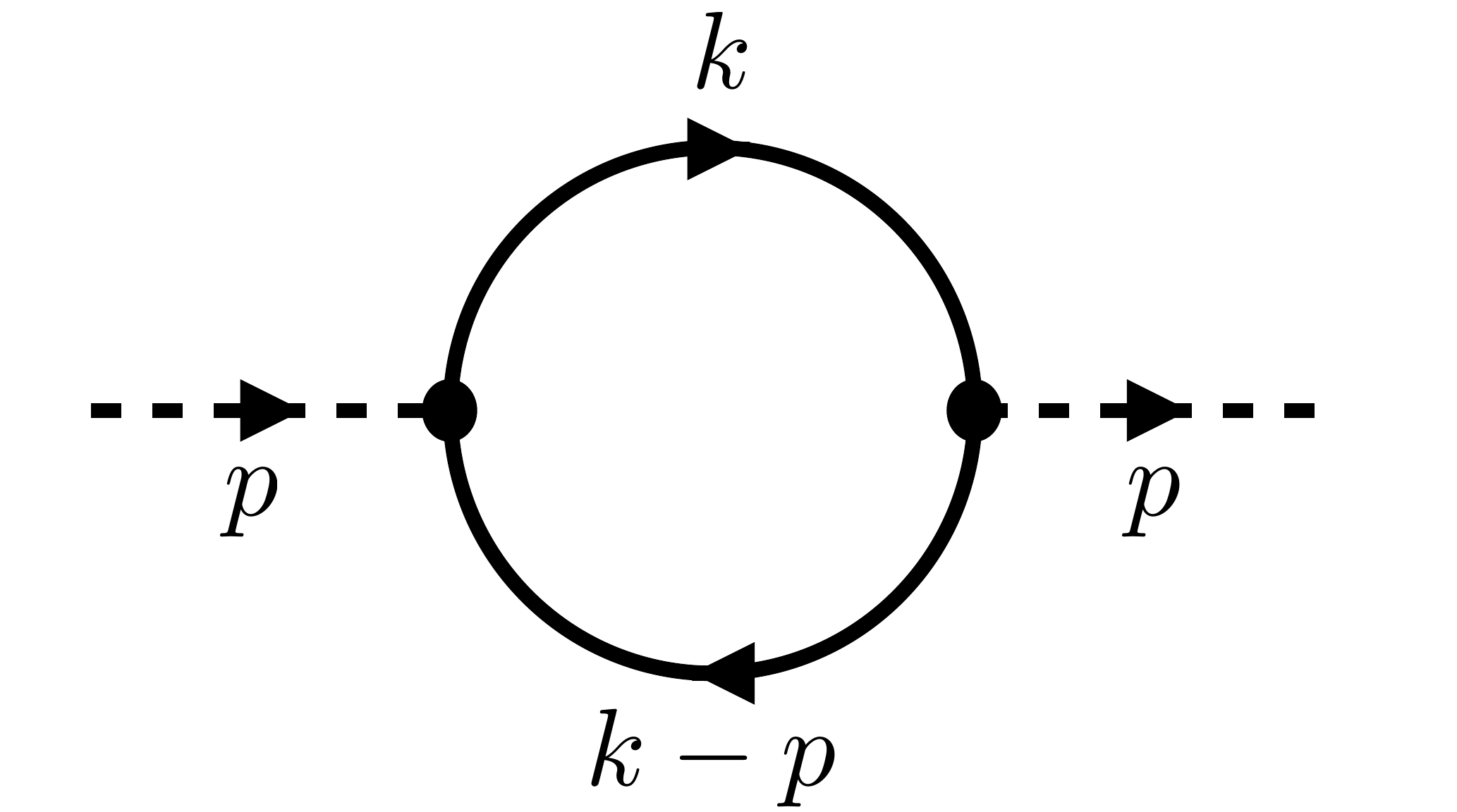}
\caption{One-loop self energy for the scalar field $\phi$ which is represented by dashed lines, whereas solid lines represent the fermion field $\psi$.}\label{loop1}
\end{figure}

The diagram of interest is given in figure \ref{loop1}. 
The LIV is introduced via a cutoff function $f$ and, according to the discussion above, we can use 
\begin{equation}\label{popo}
\frac{i}{\slashed{p}-m+i\epsilon}\rightarrow\frac{i f(\textbf{p}^2 /\Lambda^2)}{\slashed{p}-m+i\epsilon}
\end{equation}
for the fermionic propagator\footnote{Here we neglect the MDR $\Delta$ and $\tilde{\Delta}$ and introduce only the cutoff function, without loss of generality, due to the fact that the scale which appears in the MDR is assumed to be the same as the cutoff scale.} with $f(0)=1$ and $f(\infty)=0$ (i.e. UV finiteness). In this case we will write for the propagator $G(p)=f(\textbf{p}^2 /\Lambda^2) S_{0}(p)$ where $S_{0}$ is the standard fermion propagator. Moreover, note that we are using  $\textbf{p}^2 /\Lambda^2$ as argument of the LIV cutoff function $f$ instead of $|\textbf{p}|/\Lambda$ used in ref.~\cite{Collins:2004bp}. This choice is motivated by computational simplicity and clearly it does not affect the general behaviour of the final result. Nonetheless, it is worth stressing that such a choice is CPT (and even P) invariant and consequently will end up generating only CPT even terms.

In order to determine $\xi$ it is first convenient to calculate
\begin{align}\label{cab}
c_{ab}& \equiv\frac{\partial}{\partial p^{a}}\left.\frac{\partial}{\partial p^{b}}\Pi(p)\right|_{p=0}\\ \nonumber
& = -i g^2\int\frac{d^{4}k}{(2\pi)^{4}}\left. tr\left[\left(\frac{\partial}{\partial p^{a}}\frac{\partial}{\partial p^{b}}G(k-p)\right)G(k)\right]\right|_{p=0}\\ \nonumber
& = i g^2\int\frac{d^{4}k}{(2\pi)^{4}}tr\left[\frac{\partial G(k)}{\partial k^{a}}\frac{\partial G(k)}{\partial k^{b}}\right],
\end{align}
where in the last line we have used that $\partial/\partial p^{a}=-\partial/\partial k^{a},$ performed an integration by parts and discarded a boundary term in view of the asymptotic properties of $f$. Using the above expression we can now express $\xi$ as\footnote{Actually, in this expression, we could have used $c_{22}$ or $c_{33}$ instead of $c_{11}$, as they are all equal because we are assuming Lorentz breaking only in the boost, i.e. that the rotation symmetry in space is unbroken.} 
\begin{equation}\label{xi}
\xi\equiv c_{00}+c_{11}.
\end{equation}  
Note that the trace in eq.~\eqref{cab} is such that (see appendix~\ref{appA} for details)

\begin{equation}\label{tra}
tr\left[\frac{\partial G(k)}{\partial k^{a}}\frac{\partial G(k)}{\partial k^{b}}\right]= k_{a}k_{b}F+\eta_{ab}G,
\end{equation}
where $F$ and $G$ are scalar functions and $\eta_{ab}$ is the flat spacetime metric. From here it is already clear that $c_{ab}$ does not vanish only if $a=b$. Indeed, while the second term on the r.h.s. is zero when $a\neq b$, the first one vanishes, in view of its symmetry, after the integration in eq.~\eqref{cab}. 
Given eq.~\eqref{trace}, which provides an expression for the trace in eq.~\eqref{tra} when $a=b$, we can now compute the coefficients of interest. For doing this it is convenient to Wick rotate $k^{0}\rightarrow ik^{0}$ to work in Euclidean space\footnote{Wick rotating here comes without problems. Indeed the location of the poles of the integrand in the complex $k^{0}$-plane is exactly the same as in standard QFT and therefore there are no obstructions for rotating the integration contour on the imaginary axis. Alternatively, one can avoid Wick rotation and instead use the residue theorem taking care of the $i\epsilon$ terms.} and to compute first the integral over $k^{0}$, since the unknown cutoff function $f$ does not depend on it. Defining, for convenience, $$A=\textbf{k}^{2}+m^{2}$$ (where $\textbf{k}$ indicates the 3-momentum) and using eqs.~\eqref{intc1} and~\eqref{intc0} we finally arrive at
\be\label{c00}
c_{00}=g^{2}\int\frac{d^{3}k}{(2\pi)^{3}}f^{2}\left(-\frac{1}{A^{3/2}}+\frac{m^{2}}{A^{5/2}}\right),
\ee
and, after some algebra at
\be\label{c11}
c_{ii}=-g^{2}\int\frac{d^{3}k}{(2\pi)^{3}}\left[\frac{(k^{i})^{2}}{\Lambda^{2}}\left(-\frac{8(f')^{2}}{\Lambda^{2}A^{3/2}}k^{2}-4\frac{f\,f'}{A^{5/2}}(-k^{2}+2m^{2})\right)+\frac{5m^{2}f^{2}(k^{i})^{2}}{A^{7/2}}-\frac{f^{2}}{A^{3/2}}\right],
\ee
where, hereafter, $k=|\textbf{k}|$ and the argument of $f$ and its derivative $f'$ is understood to be $k^{2}/\Lambda^2$. In order to compute $\xi$ according to eq.~\eqref{xi} we exploit the spherical symmetry of the problem and integrate on the angular variables, making explicit the fact that $c_{ii}$ is actually independent of $i$. In this way we conclude that 
\be\label{ad}
\xi=-\frac{g^{2}}{2\pi^{2}}\int_{0}^{\infty}dk\,k^{2}\left[f^{2}\frac{m^{2}}{A^{7/2}}\left(\frac{5}{3}k^{2}-A\right)-\frac{8k^{4}}{3\Lambda^{4}A^{3/2}}(f')^{2}+\frac{4k^{2}(k^{2}-2m^{2})}{3\Lambda^{2}A^{5/2}}f\,f'\right].
\ee
This expression can be cast in a simpler form by introducing the dimensionless variable $y=k^{2}/\Lambda^{2}$ and the ratio $\rho=m^{2}/\Lambda^{2}$. Accordingly,
\begin{align}\label{xicollins}
\xi&=-\frac{g^{2}}{2\pi^{2}}\int_{0}^{\infty}dy\frac{\sqrt{y}}{2}\left[f^{2}\left(\frac{2}{3}y-\rho\right)\frac{\rho}{(y+\rho)^{7/2}}-\frac{8}{3}y^{2}(f')^{2}\frac{1}{(y+\rho)^{3/2}}+\frac{4}{3}y\,f\,f'\frac{y-2\rho}{(y+\rho)^{5/2}}\right]\\ \nonumber
&=-\frac{g^{2}}{2\pi^{2}}\left[-\frac{4}{3}\int_{0}^{\infty}dy\,y^{5/2}(f')^{2}\frac{1}{(y+\rho)^{3/2}}+\frac{2}{3}\int_{0}^{\infty}dy\,y^{3/2}\,f\,f'\frac{y-\rho}{(y+\rho)^{5/2}}\right],
\end{align}
where $f$ is now a function of $y$, $f'\equiv df(y)/dy$ and on the second line we have used that 
$$
\int_{0}^{\infty}dy\frac{\sqrt{y}}{2}f^{2}\left(\frac{2}{3}y-\rho\right)\frac{\rho}{(y+\rho)^{7/2}}= \frac{2}{3}\int_{0}^{\infty}dy\,\frac{y^{3/2}}{(y+\rho)^{5/2}}f\,f'\,\rho,
$$ 
which holds for the first integral on the first line of eq.~\eqref{xicollins} up to vanishing boundary terms. We are interested here in the IR percolation of LIV, i.e. in the value of $\xi$ in the formal limit $\Lambda\rightarrow\infty$, corresponding to having $\Lambda$ much larger than any mass scale $m$, which also implies $\rho\rightarrow 0$. In this limit, eq.~\eqref{xicollins} gives 
\begin{equation}
\xi=\frac{g^{2}}{6\pi^{2}}\left[1+4\int_{0}^{\infty}dy\,y\,(f'(y))^{2}\right].
\label{eq:res-collins}
\end{equation} 
Taking into account that ref.~\cite{Collins:2004bp} considers a cutoff function $\hat f$ (denoted therein simply by $f$) which depends on $x\equiv |\textbf{k}|/\Lambda$ and not on $y=x^2$ as we do here, with $\hat{f}(x)=f(x^2)$ the previous equation becomes 
\begin{equation}
\xi=\frac{g^{2}}{6\pi^{2}}\left[1+2\int_{0}^{\infty}dx\,x\,(\hat{f}'(x))^{2}\right],
\end{equation}
which indeed coincides with the result reported in eq.~(A.2) of ref.~\cite{Collins:2004bp}, after the change of notation $\hat{f}\rightarrow f$.
Let us note that, given the absence of MDR, this result can be also seen as a recovery of the well-known fact that a Lorentz invariance violating cutoff leaves a  ``LIV memory'' even when it is formally removed. 

We emphasise here that $\xi$ can be interpreted as the fractional deviation with respect to the speed of light $c=1+\mathcal{O}(g^{2})$ (assumed to equal one at tree level for convenience) in a perturbative expansion in the coupling constant (see refs.~\cite{Urrutia:2005ty,Alfaro:2004aa}), i.e.
\begin{equation}
\frac{\Delta c}{c}= \xi +\mathcal{O}(g^{4}).
\end{equation}
This relationship is valid also in the case of the fermion discussed in the next section. For this reason, hereafter we will indicate by $\Delta c$ (instead of $\xi$) the quantity representing the LIV percolation.

\section{Fermion self energy\label{section3}}

In this section we focus on the LIV percolation on the fermion and therefore we consider the fermion self-energy at one loop. We will use the previous setting  concerning the propagators. In this case we obtain some interesting results beyond corroborating the generality of the argument in ref.~\cite{Collins:2004bp} (see also ref.~\cite{Urrutia:2005ty} for a similar analysis).

\begin{figure}[tbp]
\centering
\includegraphics[trim={0 10.5cm 0 9cm},clip, scale=0.4]{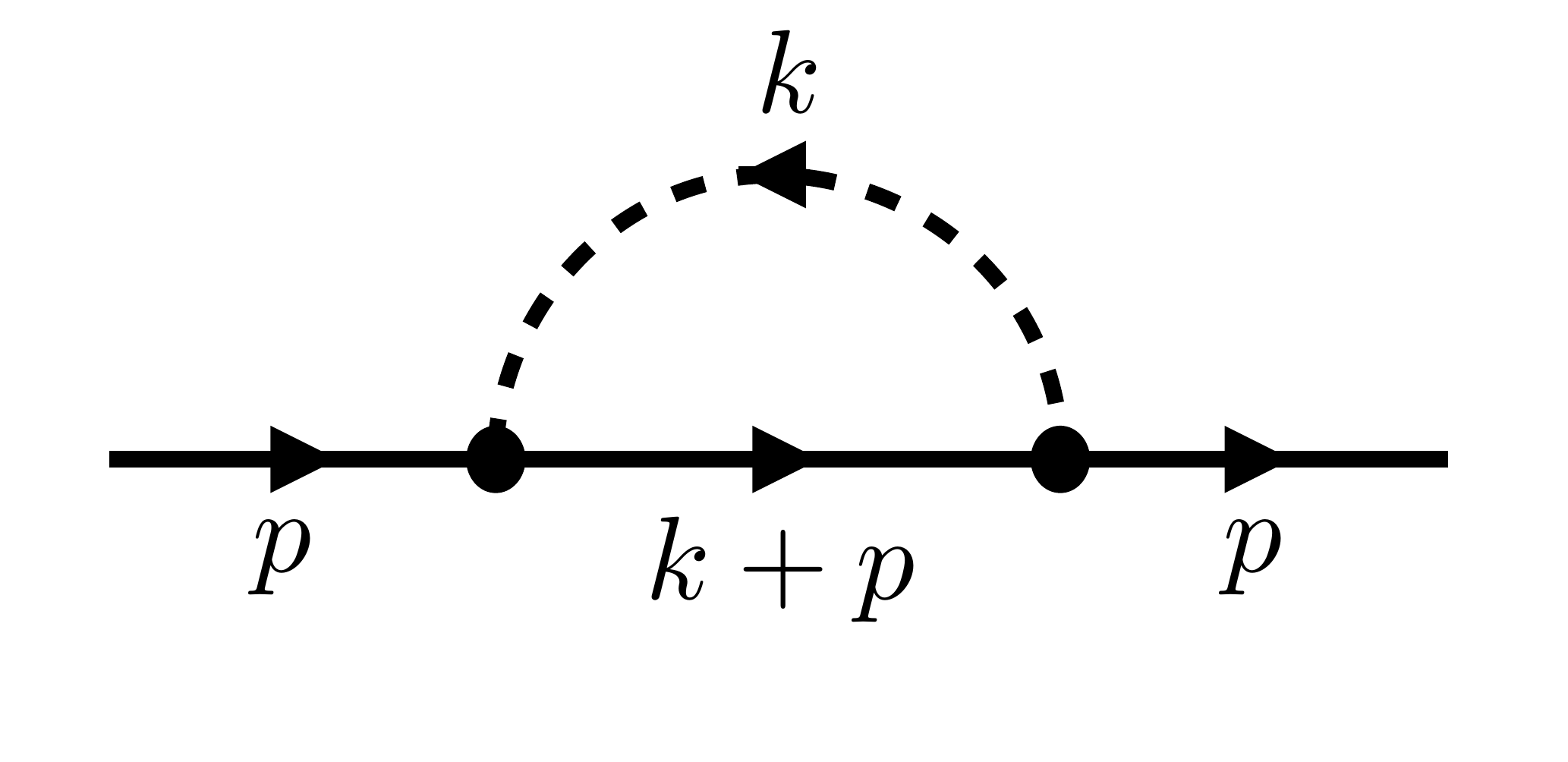}
\caption{One-loop self-energy for the fermion. The dashed line represents the scalar field $\phi$ while solid ones represent the fermion $\psi$.}\label{loop2}
\end{figure}

The relevant diagram contributing to the self-energy $\Sigma$ of the fermion is represented in figure \ref{loop2}. 
%
%
%
%
Note that, since now one fermion and one scalar are involved in the loop, we can choose which field carries the LIV, i.e.~which cutoff function $f$ or $\tilde{f}$ to introduce. 
These options give rise to different cases that we analyze separately further below. In particular, we assume in full generality that the scalar and fermion fields have not only unequal masses $m_{\phi}$ and $m_{\psi}$ but also different LIV cutoff functions $f$ and $\tilde{f}$, respectively,  with no MDR, as assumed in ref.~\cite{Collins:2004bp}. Then, we  specialise the corresponding general expressions to particular cases. 

The self-energy $\Sigma$ is given by
\be
\Sigma(p)=-ig^2 \int\frac{d^{4}k}{(2\pi)^4}\frac{\slashed{k}+\slashed{p}+m_{\psi}}{(k+p)^2-m_{\psi}^{2}}\tilde{f}\left(|\textbf{k}+\textbf{p}|^{2}/\Lambda^{2}\right)\;\frac{f\left(|\textbf{k}|^{2}/\Lambda^{2}\right)}{k^2 -m_{\phi}^2}.
\ee
This expression can be decomposed in three parts, i.e. 
\begin{equation}\label{sigma}
\Sigma=\Sigma_{1}+\Sigma_{2}+\Sigma_{3},
\end{equation}
where
\begin{align}
& \Sigma_{1}= -ig^2 \int\frac{d^{4}k}{(2\pi)^4}\frac{m_{\psi}}{(k+p)^2-m_{\psi}^{2}}\tilde{f}\left(|\textbf{k}+\textbf{p}|^{2}/\Lambda^{2}\right)\;\frac{f\left(|\textbf{k}|^{2}/\Lambda^{2}\right)}{k^2 -m_{\phi}^2}\equiv m_{\psi}\chi_{1}(p),\\
& \Sigma_{2}=-ig^2 \int\frac{d^{4}k}{(2\pi)^4}\frac{\slashed{k}}{(k+p)^2-m_{\psi}^{2}}\tilde{f}\left(|\textbf{k}+\textbf{p}|^{2}/\Lambda^{2}\right)\;\frac{f\left(|\textbf{k}|^{2}/\Lambda^{2}\right)}{k^2 -m_{\phi}^2}\equiv \chi_{2}(p),\\
& \Sigma_{3}=-ig^2 \int\frac{d^{4}k}{(2\pi)^4}\frac{\slashed{p}}{(k+p)^2-m_{\psi}^{2}}\tilde{f}\left(|\textbf{k}+\textbf{p}|^{2}/\Lambda^{2}\right)\;\frac{f\left(|\textbf{k}|^{2}/\Lambda^{2}\right)}{k^2 -m_{\phi}^2}\equiv \slashed{p}\chi_3(p).
\end{align}
The one-loop form of the fermion inverse propagator in momentum space is
\be\label{ol}
\slashed{p}-m_{\psi}+\Sigma(p)=\slashed{p}-(1-\chi_{1}(0))m_{\psi}+c_{0}\gamma^{0}p^{0}-c_{i}\gamma^{i}p^{i},
\ee
where $\gamma^{a}$ are gamma matrices and we have expanded the self energy around $p=0$, neglecting higher-order terms. In order to extract the LIV we will compute the coefficients
\be\label{coeff}
c_{a}=\frac{1}{4} \left.tr\left[\gamma^{a}\frac{\partial\Sigma}{\partial p^{a}}\right]\right|_{p=0}.
\ee
They are analogous to the $c_{aa}$ coefficients we determined in the case of the scalar; a LIV amounts to a non vanishing 
\be\label{di}
\Delta c\equiv c_{0}-c_{i},
\ee
where $i=\left\{1,2,3\right\}$ stands for a spatial index. As before, the result will be independent of the particular $i$ chosen since rotational invariance is unbroken. The fact that in standard LI QFT this quantity vanishes can be checked directly by using, e.g. dimensional regularization.
Moreover, note that the coefficients $c_{a}$ are alternatively given by 
\begin{align}
& c_{a}=\chi_{3}(0)+\frac{1}{4} tr\left.\left[\gamma^{a}\frac{\partial\chi_{2}(p)}{\partial p^{a}}\right]\right|_{p=0}=\chi_{3}(0)+\tilde{c}_{a},
\end{align}
with $a=0,1,2,3$ and therefore the violation in eq.~\eqref{di} can also be written as $\Delta c=\tilde{c}_{0}-\tilde{c}_{i}$.

The quantities of interest are given by
\begin{align}\label{cin}
c_{a}& = \frac{\partial}{\partial p^{a}}(-ig^2) \int\frac{d^{4}k}{(2\pi)^4}\frac{\eta^{ab}(k_{b}+p_{b})}{(k+p)^2-m_{\psi}^{2}}\tilde{f}\left(|\textbf{k}+\textbf{p}|^{2}/\Lambda^{2}\right)\;\left.\frac{f\left(|\textbf{k}|^{2}/\Lambda^{2}\right)}{k^2 -m_{\phi}^2}\right|_{p=0}\\ \nonumber
& =(-ig^2) \int\frac{d^{4}k}{(2\pi)^4}\left.\left\{\frac{\tilde{f}f}{(k^2 -m_{\psi}^2)(k^2 -m_{\phi}^2)}+k^{a}\left(\frac{\left.\frac{\partial \tilde{f}}{\partial p^{a}}\right|_{p=0}}{k^2 -m_{\psi}^2}-\frac{2k^{a}\eta_{aa}}{(k^2 -m_{\psi}^{2})}\tilde{f}\right)\frac{f}{k^2 -m_{\phi}^{2}}\right\}\right|_{p=0}, 
\end{align}
where in the last line the dependence of $f$ and $\tilde{f}$ on $y \equiv |\textbf{k}|^2/\Lambda^{2}$ is understood.
By using the chain rule one has
\be
\left.\frac{\partial \tilde{f}}{\partial p^{a}}\right|_{p=0}=
\begin{cases}
0 &\quad \mbox{for} \quad a = 0,\\[1mm]
\displaystyle{\frac{2k^{a}}{\Lambda^2}\tilde{f}'} &\quad \mbox{for} \quad a \neq 0
\end{cases}
\ee
where $\tilde{f}'=d\tilde{f}(y)/dy$, i.e. this term is present only for $a\neq 0$.

Now both $c_{0}$ and $c_{i}$ can be read from eq.~\eqref{cin}:
\begin{align}
& c_{0}=(-ig^2) \int\frac{d^{4}k}{(2\pi)^4}\left\{\tilde{f}f\left[\frac{1}{(k^2 -m_{\psi}^2)(k^2 -m_{\phi}^2)}-\frac{2(k^{0})^{2}}{(k^2 -m_{\psi}^2)^{2}(k^{2}-m_{\phi}^{2})}\right]\right\}, \\
& c_{i}= (-ig^2) \int\frac{d^{4}k}{(2\pi)^4}
\!\begin{aligned}[t]
&\left\{\tilde{f}f\left[\frac{1}{(k^2 -m_{\psi}^2)(k^2 -m_{\phi}^2)}+\frac{2(k^{i})^{2}}{(k^2 -m_{\psi}^2)^{2}(k^{2}-m_{\phi}^{2})}\right]\right.\\ 
& \left.+\frac{2(k^{i})^{2}}{(k^2 -m_{\psi}^2)(k^2 -m_{\phi}^2)}f \tilde{f}'\frac{1}{\Lambda^{2}}\right\},
\end{aligned}
\end{align}
and therefore $\Delta c$ from eq.~\eqref{di} is given by 
\begin{align}\label{gen}
\Delta c =&2ig^2  \int\frac{d^{4}k}{(2\pi)^4}\tilde{f}f\frac{(k^{0})^{2}+(k^{i})^{2}}{(k^2 -m_{\psi}^2)^{2}(k^{2}-m_{\phi}^{2})}\\ \nonumber
& +ig^2  \int\frac{d^{4}k}{(2\pi)^4}\frac{2(k^{i})^2/\Lambda^2}{(k^2 -m_{\psi}^2)(k^2 -m_{\phi}^2)}f\tilde{f}'\equiv \mathcal{P}+\mathcal{Q},
\end{align}
in which we emphasise the presence of two contributions $\mathcal{P}$ and $\mathcal{Q}$. In order to proceed further we can Wick rotate the integration domain in the complex $k^{0}$-plane and compute the integral over $k^{0}$ since, as in the previous section, the cutoff functions $f$ and $\tilde{f}$ are independent of it. The integration can be done using the formulas reported in appendix~\ref{appB}. Consider first term denoted by $\mathcal{P}$ in eq.~\eqref{gen}. We can simplify it by a change of variables analogous to the one in the previous section. In particular, we define $A\equiv k^{2}+m_{\psi}^{2}$, $B\equiv k^{2}+m_{\phi}^{2}$, $z=k^{2}/m_{\psi}^{2}$, $R\equiv m_{\phi}^{2}/m_{\psi}^{2}$, $\rho=m_{\psi}/\Lambda$, where $k=|\textbf{k}|$, and we get
\begin{align}\label{p}
\mathcal{P}&=-\frac{g^2}{\pi^2}\int_{0}^{\infty}dk\; k^2 \tilde{f}\left(k^{2}/\Lambda^{2}\right)f\left(k^{2}/\Lambda^{2}\right)\left\{\frac{1}{4\sqrt{A}(\sqrt{A}+\sqrt{B})^{2}}-\frac{(2\sqrt{A}+\sqrt{B})k^{2}}{12A^{3/2}(\sqrt{A}+\sqrt{B})^{2}\sqrt{B}}\right\}\\ \nonumber
& = -\frac{g^2}{\pi^2}\int_{0}^{\infty}dz \frac{\sqrt{z}}{2} \tilde{f}\left(\rho^{2}z\right)f\left(\rho^{2}z\right)\nonumber
\!\begin{aligned}[t]
&\left\{\frac{1}{4\sqrt{z+1}(\sqrt{z+1}+\sqrt{z+R})^{2}}+\right.\\ \nonumber
& \left.-\frac{(2\sqrt{z+1}+\sqrt{z+R})z}{12(z+1)^{3/2}(\sqrt{z+1}+\sqrt{z+R})^{2}\sqrt{z+R}}\right\}.
\end{aligned}
\end{align}
In order to extract the possible IR percolation we consider the limit $\rho\rightarrow 0$, with generic mass ratio $R$, exactly as we did in section \ref{dd}. Taking into account the properties of the cutoff functions $f$ and $\tilde{f}$ for vanishing arguments, i.e. $f$, $\tilde{f}\rightarrow 1$, the remaining integral gives 
\begin{equation}\label{pp}
\mathcal{P}(\rho\ll 1)=-\frac{g^{2}}{48\pi^2},
\end{equation}
which turns out to be independent of the mass ratio $R$. 

The second term in eq.~\eqref{gen}, i.e. $\mathcal{Q}$, can now be calculated after performing a Wick rotation followed by the integration over $k^{0}$ and by the same change of variables as above
\begin{align}\label{q}
\mathcal{Q}& =-g^2  \int\frac{d^{3}k}{(2\pi)^3}f\tilde{f}'\frac{2(k^{i})^{2}}{\Lambda^{2}}\int\frac{dk^{0}}{(2\pi)}\frac{1}{((k^{0})^{2}+A)((k^{0})^{2}+B)}\\ \nonumber
& = -g^2  \int\frac{d^{3}k}{(2\pi)^3}f\tilde{f}'\frac{(k^{i})^{2}}{\Lambda^{2}}\frac{1}{2}\frac{1}{A\sqrt{B}+B\sqrt{A}}\\ \nonumber
& = -\frac{g^2}{12\pi^2}\int_{0}^{\infty}dz \sqrt{z}\rho^2 z f(\rho^2 z)\tilde{f}'(\rho^2 z)\frac{1}{(z+1)\sqrt{z+R}+(z+R)\sqrt{z+1}},
\end{align}
where $\rho$, $R$ and $z$ are given right before eq.~\eqref{p}.
This expression can be simplified as
\be\label{qq}
\mathcal{Q}=-\frac{g^2}{12\pi^2}\int_{0}^{\infty}dz \sqrt{z} z f(\rho^2 z)\frac{d}{dz}\left[\tilde{f}(\rho^2 z)\right]\frac{1}{(z+1)\sqrt{z+R}+(z+R)\sqrt{z+1}}.
\ee
In order to proceed further with the calculation of $\mathcal{Q}$ we need to consider below specific choices for the functions $f$ and $\tilde{f}$. Note, however, that having generically $\mathcal{P},\mathcal{Q}\neq 0$ suggests that the percolation will be unsuppressed, i.e. that $\Delta c\neq 0$ unless a cancellation occurs.

\subsection{Particles with equal masses ($R=1$) and same violation ($f=\tilde{f}$)}

As a first simplification, we assume that both the masses $m_{\phi}$ and $m_{\psi}$ and the LIV cutoff functions $f$ and $\tilde{f}$ of the fields are equal. 
Though a priori there is no reason for the latter assumption, one could argue that QG affects both fermionic and bosonic fields in exactly the same way, hence suggesting $f=\tilde{f}$. In this case we can use that $f\cdot f'(z)=1/2\;d(f^2)/dz$, and an integration by parts of eq.~\eqref{qq} yields
\be
\mathcal{Q}=\frac{g^2}{12\pi^2}\int_{0}^{\infty}dz \frac{1}{2} f^2(\rho^{2} z) \frac{d}{dz}\left(\frac{\sqrt{z} z}{2(z+1)^{3/2}}\right),
\ee
where the contribution of the boundary terms stemming from the integration vanishes due to the behaviour of the function $f$, while the remaining part can be integrated and, in the limit $\rho\rightarrow 0$ (which implies $f\rightarrow 1$), it gives $g^2/(48\pi^2)$ which is equal and opposite to $\mathcal{P}$ in eq.~\eqref{p} and therefore (see eq.~\eqref{gen})
\begin{equation}
\label{eq:no-viol}
\Delta c (\rho \ll 1)=0. 
\end{equation}
Remarkably, in this case, the percolation on the fermion is absent albeit the calculation of section \ref{sec2} shows that this is not the case for the scalar. In order to understand how general this fact is, we consider below the case in which the two particles still carry the same LIV ($f=\tilde{f}$) but have different masses.

\subsection{Particles with different masses ($R\neq 1$) and same violation ($f=\tilde{f}$)}\label{32}

Under the assumption $f=\tilde{f}$, eq.~\eqref{q} can be integrated by parts, as done above. The associated boundary terms vanish and one is left with
\begin{equation}
\mathcal{Q}= +\frac{g^2}{12\pi^2}\int_{0}^{\infty}dz \frac{1}{2} f^2(\rho^{2} z) \frac{d}{dz}\left(\frac{\sqrt{z} z}{(z+1)\sqrt{z+R}+(z+R)\sqrt{z+1}}\right).
\end{equation}
In spite of having $R\neq 1$, this integral still gives $g^2/(48\pi^2)$ for $\rho\rightarrow 0$ (i.e. $m_{\psi}\ll\Lambda$) and therefore $\Delta c$ vanishes as in eq.~\eqref{eq:no-viol}, independently of the values of the masses $m_{\phi,\psi}$. This case is more interesting than the previous one $m_\phi = m_\psi$ as it suggests that, with an heavy scalar field $m_{\phi}\gg m_{\psi}$, the low-energy physics of the fermion field will not be affected by the LIV unsuppressed percolation on the scalar computed in section \ref{sec2}, because no percolation is present on the fermion and indeed the scalar can be integrated out. As such, it would be interesting to check whether this scenario could be extended to the SM and Higgs field.

\subsection{General case: $R\neq 1$ and $f\neq\tilde{f}$}
Integrating by parts eq.~\eqref{qq} one finds that:  
\begin{equation}
\mathcal{Q}=\frac{g^2}{12\pi^2}\int_{0}^{\infty}dz \tilde{f}\left(\rho^{2}z\right) \frac{d}{dz}\left[\frac{\sqrt{z} z f\left(\rho^{2}z\right)}{(z+1)\sqrt{z+R}+(z+R)\sqrt{z+1}}\right],
\end{equation}
where as before, the boundary terms vanish because of the properties of $f$ and $\tilde{f}$.
In this case the result for $\rho\rightarrow 0$ is $\mathcal{Q}(\rho\ll 1)=g^{2}/(24\pi^{2})$ and therefore
\begin{equation}
\label{eq:vi-coll}
\Delta c (\rho \ll 1)=\frac{g^{2}}{48\pi^2},
\end{equation}
independently of the specific form of $\tilde{f}$ and $f$ and on the value of $R$. 

\subsection{Violation only on the scalar field ($\tilde{f}=1$)}\label{only}
Finally, we want to specialise eqs.~\eqref{pro1} and~\eqref{pro2}  to the case $\tilde{f}=1$, in which only the scalar propagator carries the LIV. 
(Note that the corresponding expression for $\Delta c$ cannot be derived directly from the ones discussed above, as they assume that $\tilde f$ vanishes for large values of its argument, which is not the case here.)
Starting from eq.~\eqref{gen} we see that $\mathcal{Q}=0$ and therefore the only contribution to $\Delta c$ is due to $\mathcal{P}$ in eq.~\eqref{p} with $\tilde{f}=1$. Accordingly, the result of the integration in the IR limit is given by 
\begin{equation}
\label{eq:vi-new}
\Delta c (\rho \ll 1)=-\frac{g^{2}}{48\pi^2}.
\end{equation}
This shows that again there is an unsuppressed percolation as in ref.~\cite{Collins:2004bp} (and in accordance with ref.~\cite{Urrutia:2005ty}). 

\section{Separation of scales}
\label{sec4}

Now that we have reviewed and extended the results presented in the literature we are going to show how the introduction of a LI cutoff, in addition to a LIV modified dispersion relation, can hinder the percolation. As anticipated in the introduction, the quest for a mechanism able to prevent the IR percolation of LIV is not new (see, e.g. refs.~\cite{Liberati:2013xla,Myers:2004ge}). In particular, there were proposals based on having supersymmetry as a custodial symmetry.
What we are going to show in the following is somehow related to this custodial symmetries protection mechanism. The idea is that if there is a separation between the EFT validity scale $\Lambda$ (i.e. the scale of possible new physics beyond the SM) and the LIV scale\footnote{$M$ can be assumed to coincide with the Planck scale due to the fact that we can expect LIV coming from the scale at which our concept of spacetime as a pseudo-Riemannian manifold is questionable together with the associated symmetries.} $M$, with $\Lambda<M$, then the IR percolation is suppressed  by a power of the ratio $\Lambda/M$ which eventually controls its magnitude. This result, from the EFT perspective, is rather natural because the introduction of a new mass scale $\Lambda$ gives the possibility to have a (small) dimensionless ratio. 

Note that, we are not arguing here that the one discussed below is a protection mechanism which works for the entire Standard Model (SM) nor that there will be room in the SM for such a large scale separation to suppress low-energy LIV in a way which complies with the strong bounds coming from observational data. What we want to emphasise, via a toy-model computation, is that the separation of scales could be one, or part of a, solution to the naturalness problem of LIV and in this way show the validity of some heuristic ideas presented in the literature.

In order to investigate how this separation of scales hinders the percolation, we consider again the model defined by eq.~\eqref{lag}. In particular we introduce a LI cutoff (as explained below) called $\Lambda$ that represents the scale of validity of the EFT description and we consider 
the case in which the LIV is carried only by the scalar field and is encoded in a MDR through the scale $M$ possibly associated with some QG scenario. We will then be interested in the case in which both scales are larger than the masses of the particles in the problem and analyze the effect of the separation of scales on the LIV percolation. To be concrete, we choose some particular forms of the MDR in order to be able to carry out numerically the calculation of the resulting percolation. However, we will argue that the result is largely independent of these choices. Note that, the case under study is physically interesting since the one-loop scalar self-energy does not receive a contribution from the fermion loop as we assume that the fermion field is LI at the tree-level. Accordingly, we have to consider only the fermion self-energy depicted in figure~\ref{loop2} . 

The calculation of $\Delta c$ is similar to the one presented in section~\ref{section3}, with the difference that we want to introduce here a new LI cutoff $\Lambda$ that represents the scale of validity of the EFT, i.e. the scale of new, Lorentz-invariant physics. This is done in two different ways: The first is a LI sharp cutoff on the 4-momentum, which does not break LI although it clearly breaks Poincar\'e invariance.  
The second is a LI cutoff introduced via a smooth non-local function which can be thought of as deriving from a fundamentally non-local theory in which the non-locality improves the UV behaviour of the theory (see, however, section~\ref{nlc} for additional comments). Noticeably, these non-local features have been suggested in the quantum gravity literature as a possible low-energy signature of the microscopic nature of spacetime. In both cases we will conveniently implement the cutoff only after rotation in Euclidean space (see also the discussion in section~\ref{nlc}). 

Before considering the presence of two different scales $\Lambda$ and $M$ let us make a remark on the case with a single scale investigated in the previous sections. Note that the MDR itself can serve as a LIV cutoff at the scale set by $M$. Indeed, it can be shown (see appendix~\ref{appF}) that upon increasing $M$ we recover a finite percolation as predicted in ref.~\cite{Collins:2004bp} but with a value of $\Delta c$ which depends on the specific form of the MDR used. In particular, considering a MDR  for the scalar field of the form given by eq.~\eqref{pro2} with $\tilde{f}=1$ and  
\begin{equation}
\label{MDR}
\Delta(|\textbf{k}|,M)\equiv -|\textbf{k}|^{2}\left(\frac{|\textbf{k}|^{2}}{M^{2}}\right)^{n},
\end{equation}
with $n>0$, which is typically encountered in QG phenomenology, $\Delta c$ approaches 
\begin{equation}
\label{asi} 
\Delta c(m_{\phi,\psi}/M \ll 1) = -\frac{g^{2}}{48\pi^{2}} \frac{n+1}{n},
\end{equation}
for large values of $M$ compared to the mass scales in the problem, see appendix~\ref{appF} and compare with eq.~\eqref{eq:vi-new}.
From this expression it is clear that the percolation is always larger in modulus (by a numerical factor) than the one found in the absence of MDR but using a LIV cutoff on the spatial momenta as in ref.~\cite{Collins:2004bp}. Indeed, the present case in which the MDR effectively introduces a LIV regularization in the UV of the spatial part of the integral, is intrinsically different from the one considered in ref.~\cite{Collins:2004bp} given that there (and in sections \ref{sec2} and \ref{section3} here) the cutoff function is such that it renders the theory UV finite independently of the order of the radiative corrections considered while the MDR cannot achieve  this in general.

\subsection{Sharp LI cutoff}\label{sharp}\label{sec41}
First, we consider the case of a sharp cutoff $\Lambda$ on the four momentum. We work in the Euclidean space where this cutoff has the effect of restricting the integration inside a sphere of radius $\Lambda$. The computation of $\Delta c$ follows the same lines as in section \ref{section3} and, in fact, we arrive at an equation similar to eq.~\eqref{gen}:
\begin{equation}\label{hard}
\Delta c =2ig^2  \int_{\Lambda}\frac{d^{4}k}{(2\pi)^4}\frac{(k^{0})^{2}+(k^{i})^{2}}{(k^2 -m_{\psi}^2)^{2}\left(k^{2}-m_{\phi}^{2}+\Delta(|\textbf{k}|,M)\right)},
\end{equation}
where $\textbf{k}=\left\{k^{1},k^{2},k^{3}\right\}$ is the 3-momentum and $\Delta$ is given by eq.~\eqref{MDR}. This expression is still in Minkowski space and the subscript $\Lambda$ in the integral indicates that a sharp cutoff is implemented, i.e. that the domain of integration is restricted as specified further below. Here the LIV is entirely introduced by the MDR of the scalar field and we use $M$ as the LIV scale while $\Lambda$ as the LI cutoff scale. 
Now we perform a Wick rotation\footnote{It is easy to check that the locations of the poles of eq.~\eqref{hard} in the complex $k^{0}$-plane are such that a rotation of the integration contour is possible.} and use 4D spherical coordinates in order to evaluate the integrals. Accordingly, eq.~\eqref{hard} becomes

\begin{align}\label{hardsph} 
\Delta c &= -2g^{2}\int_{\Lambda} \frac{d^{4}k}{(2 \pi )^4}\frac{(k^{0})^{2}-(k^{1})^{2}}{\left(k^{2}+m_{\psi }^{2}\right)^{2} \left(k^{2}+\textbf{k}^{2}\left(\frac{\textbf{k}^{2}}{M^2}\right)^{n}+m_{\phi}^{2}\right)}\\ \nonumber
&= -\frac{g^{2}}{2 \pi^3}\int _{0}^{\Lambda }dk\,\frac{k^5}{\left(m_{\psi }^2+k^2\right)^{2}} \int _{0}^{\pi }d\phi\frac{\cos^{2}\phi-(\sin^{2}\phi) /3}{\left(k^{2}+m_{\phi}^{2}+
{\frac{\textstyle k^{2+2n}\sin^{2+2n}\phi}{\textstyle M^{2n}}}\right)} \sin^{2}\phi,
\end{align}
where $k$ stand for the modulus of the 4-momentum in the Euclidean space\footnote{In the numerator of the integrand of this equation, $(k^{0})^{2}-(k^{1})^{2}$ corresponds to considering $i=1$ in eq.~\eqref{di}; however, the result is clearly independent of the choice of $i\in\left\{1,2,3\right\}$.}. Note that the sharp cutoff is implemented in the Euclidean space by restricting the integration to Euclidean momenta with modulus $k\leq\Lambda$. 
In particular after the rescaling $ k=\Lambda z$ we find 
\begin{align}\label{hardresc} 
\Delta c &= -\frac{g^{2}}{2 \pi^3}\int_{0}^{1} dz\,\frac{z^{5}}{(z^{2}+R_{\psi})^{2}}\int _{0}^{\pi }d\phi\,\sin^{2}\phi\frac{\cos^{2}\phi-\left(\sin^{2}\phi\right) /3}{z^{2}+R_{\phi}+
\lambda^{n} z^{2+2n}\sin^{2+2n}\phi},
\end{align}
where $\lambda\equiv\Lambda^{2}/M^{2}$ and $R_{\phi,\psi}=m^{2}_{\phi,\psi}/\Lambda^{2}$. These variables are particularly convenient for studying the large-scale separation regime $m_{\phi,\psi}\ll\Lambda\ll M$ which simply amounts at imposing $R_{\phi,\psi}\ll 1$ and $\lambda\ll 1$.
In this regime the denominator in eq.~\eqref{hardresc} can be expanded around $\lambda=0$ 
\begin{equation}\label{expan}
\frac{1}{z^{2}+R_{\phi}+\lambda^{n} z^{2+2n}\sin^{2+2n}\phi}=\frac{1}{z^{2}+R_{\phi}}\left[1-\frac{\lambda^{n} z^{2+2n}\sin^{2+2n}\phi}{z^{2}+R_{\phi}}+\mathcal{O}(\lambda^{2n})\right].
\end{equation}
Plugging this expansion in eq.~\eqref{hardresc}, performing the angular integration and noting that the latter vanishes at the zeroth order we have
\begin{equation}\label{new1}
\Delta c=-\frac{g^{2}}{2\pi^{3}}\lambda^{n}\frac{(1+n)\sqrt{\pi}\Gamma\left(n+5/2\right)}{3\Gamma(4+n)}\int_{0}^{1}\frac{dy}{2}\frac{y^{3+n}}{(y +R_{\psi})^2 (y +R_{\phi})^2}+\mathcal{O}(\lambda^{2n}),
\end{equation}
where we introduced $y=z^{2}$ (see appendix~\ref{appE} for details on the $y$-integration in eq.~\eqref{new1}). This equation clearly shows that the violation $\Delta c\propto\left(\Lambda/M\right)^{2n}$ is suppressed whenever $M\gg\Lambda$. The actual degree of suppression depends on the separation between the LIV scale $M$ (which can be identified with the Planck scale) and the scale $\Lambda$ of the LI new physics as well as on the specific form of the MDR, i.e. on $n$. 

\begin{figure}[tbp]
\centering
\includegraphics[scale=0.6]{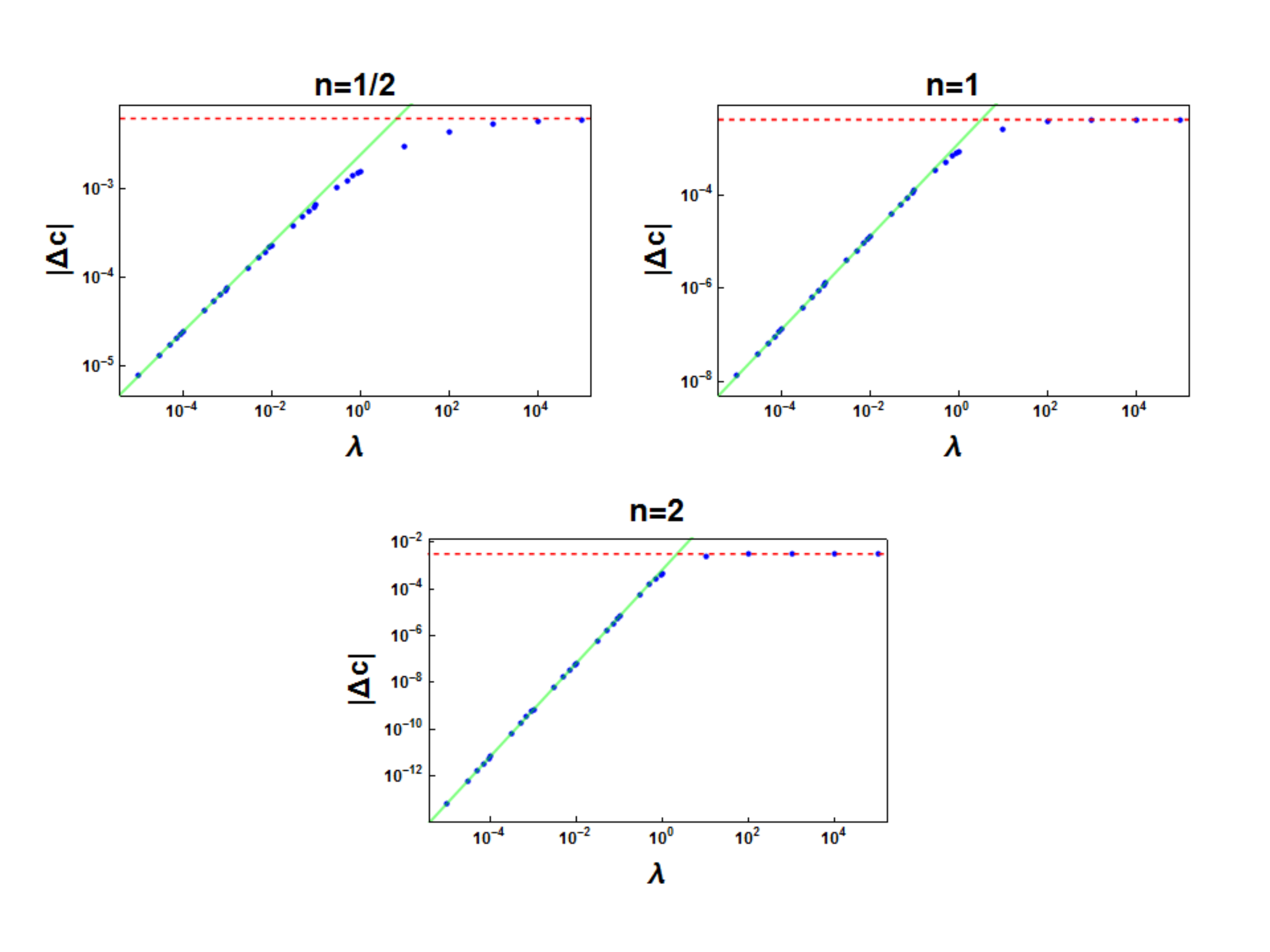}
\caption{Dependence of $|\Delta c|$ on $\lambda\equiv\left(\Lambda/M\right)^{2}$ in the case of a sharp cutoff (see eq.~\eqref{hardresc}) with $n=1/2, 1, 2$, $R_{\phi}=R_{\psi}=10^{-12}$ and $g=1$. Symbols correspond to numerical data while the solid lines correspond to eq.~\eqref{new1} which well describes the violation at small values of $\lambda$. The dashed lines correspond to eq.~\eqref{asi}, i.e. to the values approached by $|\Delta c|$  for $\lambda\gg 1$. The violation is suppressed whenever $M\gg\Lambda$, i.e. $\lambda\ll 1$, which is exactly the separation of scales invoked various times in the literature. This behaviour agrees with what is heuristically expected on the basis of the physical intuition discussed in the main text.}\label{hard1}
\end{figure}

In particular, the suppression increases upon increasing the mass dimension of the LIV operators responsible for the MDR as the algebraic dependence on the small ratio $\Lambda/M$ has the same power as the one with which $M$ appears in the MDR. This means that, as expected, the percolation is weaker for LIV coming from mass dimension 8 operators compared to the one due to mass dimension 5 operators. Moreover,  eq.~\eqref{new1} clearly shows that $\Delta c$ (up to first order in $\lambda^{n}$) is symmetric with respect to the exchange of $R_{\psi}$ and $R_{\phi}$; the limit in which both fields are massless is finite and the dependence of $\Delta c$ on $R_{\psi,\phi}$ is rather weak at least as long as $m_{\psi,\phi} \ll \Lambda$, $M$.

Finally, we analyzed numerically also the regime $M\ll\Lambda$, i.e. $\lambda\gg 1$. Note that in order to do so the change of variable $k=M z$ is more convenient instead of the one done right before eq.~\eqref{hardresc}. The finite unsuppressed percolation in this case approaches the  value we already found in the case without the LI cutoff, i.e. eq.~\eqref{asi} (see the discussion in appendix~\ref{appF}). The numerical results for $n=1/2,\;1$ and $2$ are shown in figure \ref{hard1}, in which we assume $R_{\psi}=R_{\phi}=10^{-12}$ (i.e. equal masses).  

\subsection{Smooth LI cutoff}\label{nlc}
We close this section by considering an alternative way to introduce a LI cutoff based on non-local theories. By non-local we mean here that the kinetic term of one of the fields contains also a pseudo-differential operator of infinite order, i.e. an infinite number of spacetime derivatives acting on the field. This kind of theories are potentially unstable~\cite{Woodard:2015zca, Eliezer:1989cr},  but the Ostrogradski theorem does not apply straightforwardly. Hence they might be stable and therefore for simplicity we assume this to be the case here; moreover various works argue that they can have a better behaviour in the UV with respect to standard QFTs, while preserving the low-energy limit (see refs.~\cite{Kleppe:1991rv,Joglekar:2006ee} and also refs.~\cite{Modesto:2015lna,Tomboulis:2015gfa} and references therein). Moreover, some form of non-locality seems to be a common feature to different approaches to quantum gravity, although each theory has its own 
peculiarity~\cite{Giddings:2006vu,Calcagni:2013eua,Markopoulou:2007ha,Sorkin2006,Belenchia:2014fda}. More importantly, non-local theories can be Poincar\'e invariant while introducing a UV cutoff which can account for spacetime discreteness. In addition, it seems that a growing amount of evidence is accumulating in favor of the existence of a relationship between LI spacetime discreteness and non-locality see, e.g. refs.~\cite{Sorkin2006,Gambini:2014kba,Dowker:2003hb}.
\begin{figure}[tbp]
\centering
\includegraphics[scale=0.6]{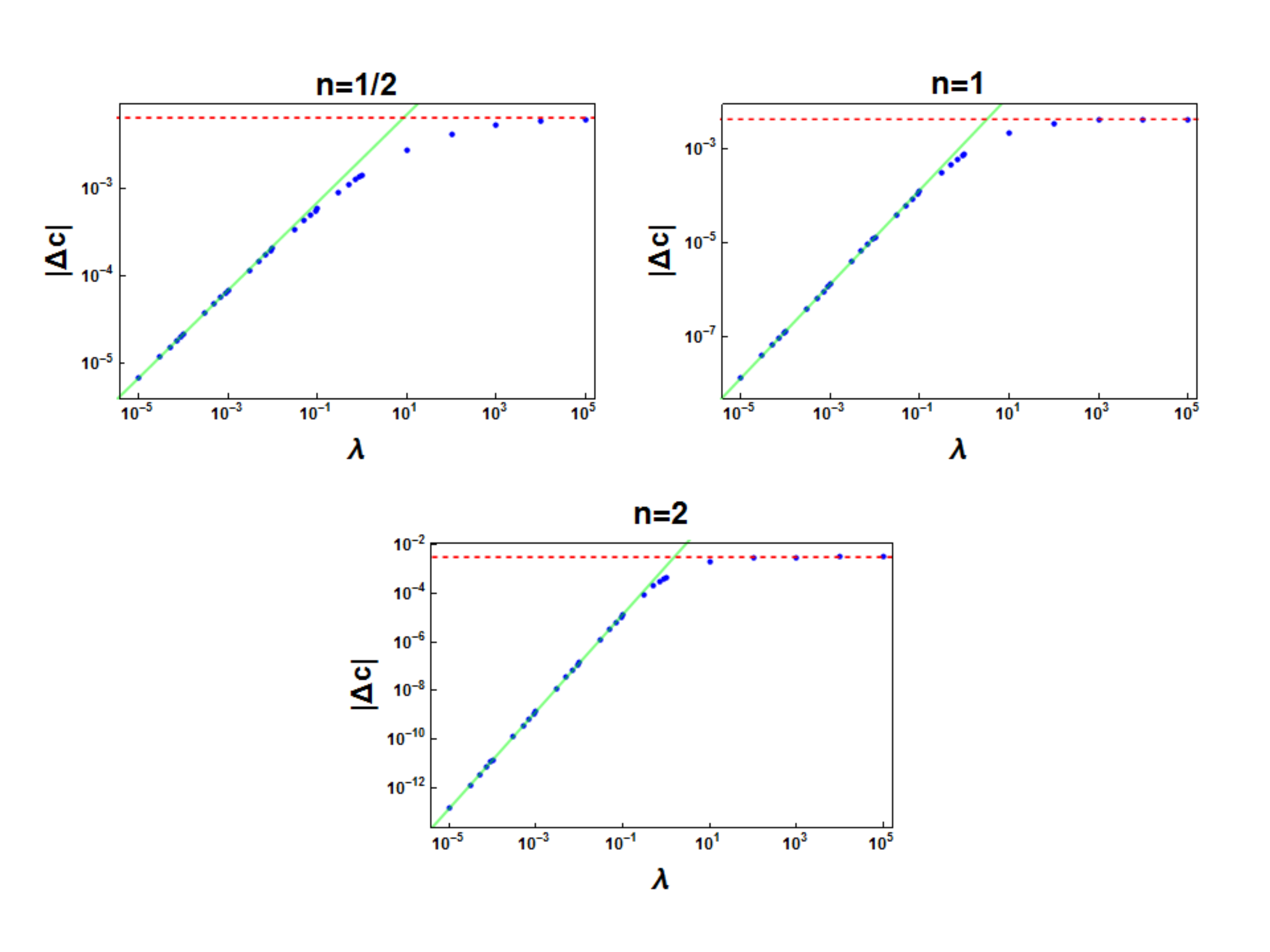}
\caption{
Dependence of $|\Delta c|$ on $\lambda\equiv\left(\Lambda/M\right)^{2}$ (see eq.~\eqref{nonlocal}) for the same values of parameters as in figure \ref{hard1}, i.e. $n=1/2, 1, 2$, $R_{\phi}=R_{\psi}=10^{-12}$ and $g=1$ but with a LI Gaussian cutoff function. Symbols correspond to numerical data while the solid lines correspond to the theoretical prediction $\Delta c\propto (\Lambda/M)^{2n}$ coming from eq.~\eqref{new1} after the replacement in eq.~\eqref{change}. The dashed lines correspond to eq.~\eqref{asi}, i.e. to the values approached by $|\Delta c|$ for $\lambda\gg 1$. The qualitative features of these curves are the same as in figure \ref{hard1}.}\label{nlcut}
\end{figure}
The possibility of such a LI regulator for QFT is also considered in ref.~\cite{Collins2006} as a way out from the apparent unavoidable link between granularity of spacetime and LIV. However, in that work it was claimed that these theories suffer from causality violations and, as such, they are not viable. Actually, the very same refs.~\cite{Jain:2003xs,Joglekar:2006ee} discussed in ref.~\cite{Collins2006} not only show that there are causality violations but also that they arise only at high energies (comparable with the non-locality scale), where the theory, if interpreted as an effective one, should anyhow cease to be valid. In this sense we shall take this smooth cutoff as not being fundamental but simply assume it to be a consequence of an EFT type description of a more fundamental theory. It is easy to see that the introduction of a smooth cutoff function $f$ in Euclidean space\footnote{Actually, the smooth cutoff has to be enforced after the rotation in the Euclidean space since the loop integrals make sense only there, see the discussion in refs.~\cite{Saini:1997ad,Evens:1990wf}} is tantamount to replacing eq.~\eqref{hardsph} by  
\begin{align}\label{nonlocal}
\Delta c &= -\frac{g^{2}}{2 \pi^3}\int _{0}^{\infty}dk\,\frac{k^5 f\left(-k^{2}/\Lambda^{2}\right)}{\left(m_{\psi }^2+k^2\right)^{2}} \int _{0}^{\pi }d\phi\frac{\cos^{2}\phi-\left(\sin^{2}\phi\right) /3 }{\left(k^{2}+m_{\phi}^{2}+
\frac{k^{2+2n}\sin^{2+2n}\phi}{M^{2n}}\right)} \sin^{2}\phi ,
\end{align}
Note that contrary to sections \ref{sec2} and \ref{section3} here $f$ depends on both the timelike and spacelike part of the 4-momentum in a LI way\footnote{See ref.~\cite{Sudarsky:2005ab} for a case in which a cutoff function dependent on both energy and momenta is introduced, even if in a way which still violates LI.} 

The behaviour of $\Delta c$  for large separation of scale $M\gg\Lambda$ can be obtained again using the expansion in eq.~\eqref{expan}, the only difference with respect to eq.~\eqref{new1} being in the integration on $y$ which now runs up to infinity and is regulated by the cutoff function, i.e.
\begin{equation}\label{change}
\int_{0}^{1}\frac{dy}{2}\frac{y^{3+n}}{(y +R_{\psi})^2 (y +R_{\phi})^2}\longrightarrow \int_{0}^{\infty}\frac{dy}{2}\frac{y^{3+n}f(-y)}{(y +R_{\psi})^2 (y +R_{\phi})^2}.
\end{equation}
We have calculated numerically eq.~\eqref{nonlocal} after the suitable rescaling of the coordinates already introduced after eq.~\eqref{hardresc}. The results for $n=1/2, 1, 2$ and the cutoff function $f(-x)=e^{-x^{2}}$ confirm those of the previous paragraph and are reported in figure \ref{nlcut}.  The violation is suppressed as $\Delta c\propto \left(\Lambda/M\right)^{2n}$, while the asymptotic values for $\Lambda\gg M$ agree with eq.~\eqref{asi} as demonstrated in appendix~\ref{appF}.


\section{Dissipation and LIV naturalness}
\label{sec5}

In the previous section we focussed on the standard picture in which the dispersion relation (in our case of the scalar field) is modified by some effects of LIV physics motivated by QG scenarios. However, general arguments~\cite{Parentani:2007uq} show that if LIV emerges dynamically from UV kinetic terms or interactions with heavy fields which are traced out in the low-energy description, then dissipative effects will also unavoidably arise. In this section we therefore focus on their influence on the percolation of LIV. Note that dissipation does not necessarily spoil unitarity, since it can emerge from bona-fide Hamiltonian models after tracing out degrees of freedom. Following these lines, ref.~\cite{Liberati:2013usa} investigated the phenomenology of dissipative effects in fields propagation, showing that strong constraints can be cast on the lower-order transport coefficients while higher-order ones are basically unconstrained. A possible percolation of higher-order dissipative terms can hence be considered an opportunity for strengthening current constraints and a motivation for our investigation. A viable model will anyhow require dissipation not to percolate strongly given the currently available constraints. In this sense we shall explore also in this case the effectiveness of a protection mechanism based on a separation of scales. 

In order to study the percolation of LIV in the presence of dissipation in the scalar field, we first address the validity of the argument of ref.~\cite{Collins:2004bp}  when there is only one relevant cutoff scale in the problem. Then we show how the picture changes when we introduce one additional scale, having the same physical interpretation as the one discussed in section \ref{sec4}. We discuss this issue both with a sharp and a smooth  LI cutoff (in the same fashion as in section \ref{sec4}).

\subsection{The general setting}
For concreteness, we consider again the model in eq.~\eqref{lag} in which dissipation affects only the scalar field. Following ref.~\cite{Parentani:2007uq}, its propagator is given by

\be\label{prodis}
G_{F}(\omega,k)=\frac{i}{\omega^{2}-k^{2}-m_{\phi}^{2}+i|\omega|k^{2+2n}/M^{1+2n}},
\ee
where $\omega$ indicates the time-like component of the momentum and $k$ the modulus of the space-like components. The parameter $n\geq 0$ is introduced for later convenience and $M$ is a mass scale. This expression can be derived from dynamical models in which heavy degrees of freedom have been traced out (see, e.g. sections 2.3 and 2.4 of ref.~\cite{Parentani:2007uq}). Note, in addition, that the dispersion relation associated with this propagator is reminiscent of the one of dissipative fluids as reported, e.g. in ref.~\cite{Liberati:2013usa}. In particular, by varying the value of $n$ we can change the first dissipative (and dispersive) term which appears in the dispersion relation (in the same way as in ref.~\cite{Liberati:2013usa}).  In analogy with the dispersive case discussed in section \ref{sec4}, we expect $\Delta c$ to depend algebraically on $\Lambda/M$ to some power which increases upon increasing $n$. 

The calculation of $\Delta c$ follows the same lines as in section \ref{section3}, leading to an expression similar to eq.~\eqref{gen}:
\begin{equation}
\Delta c =ig^2 8\pi \int \frac{dk}{(2\pi)^3}\,k^{2}\int_{-\infty}^{\infty} \frac{d\omega}{2\pi} \frac{\omega^{2}+k^{2}/3}{(\omega^2 - k^{2}-m_{\psi}^2+i\epsilon)^{2}\left(\omega^2 - k^{2}-m_{\phi}^2+i\epsilon+i|\omega|k^{2+2n}/M^{1+2n}\right)},
\end{equation}
where, taking into account the rotational symmetry, we have done the angular integration in the three spatial directions. Note that this expression refers still to Minkowski space while we do not specify, for the moment, the cutoff function that we are using. Due to the form of the dissipative term in the denominator it is not straightforward to perform a Wick rotation towards the Euclidean space. However, it is possible to show (both numerically and analytically) that $\Delta c$ given above  actually takes real values and the integral can be performed in Euclidean space. In order to do this, we can split the $\omega$ integration into two parts and use the symmetry of the argument for $\omega\rightarrow-\omega$ to write

\begin{equation}\label{syin}
\Delta c =2ig^2 8\pi \int \frac{dk}{(2\pi)^3}\,k^{2}\int_{0}^{\infty} \frac{d\omega}{2\pi} \frac{\omega^{2}+k^{2}/3}{(\omega^2 - k^{2}-m_{\psi}^2+i\epsilon)^{2}\left(\omega^2 - k^{2}-m_{\phi}^2+i\epsilon+i\omega k^{2+2n}/M^{1+2n}\right)}.
\end{equation}
Now it is clear that there are no poles in the first quadrant of the complex $\omega$-plane. Accordingly, the integral done along the path which includes the positive real $\omega$-axis up to a certain arbitrarly large value $\Omega$, the quarter of circumference of radius $\Omega$ centered in the origin $O$ of the plane and contained in its first quadrant and finally the path from $i\Omega$ to $O$ along the imaginary axis, gives zero. Noting that the integral along the quarter of circle vanishes as $\Omega\rightarrow\infty,$ we conclude that the integral along the real axis equals the one along the imaginary axis, i.e. its Wick rotation. As a result (see appendix~\ref{appD} for more details).
  
\begin{align}\label{polar}
\Delta c &=-g^2 16\pi \int \frac{dk}{(2\pi)^3}\,k^{2}\int_{0}^{\infty} \frac{d\omega}{2\pi} \frac{\omega^{2}-k^{2}/3}{(+\omega^2 + k^{2}+m_{\psi}^2)^{2}\left(\omega^2 + k^{2}+m_{\phi}^2+\omega k^{2+2n}/M^{1+2n}\right)}\\ \nonumber
&= -\frac{g^{2}}{\pi^{3}} \int d\rho\frac{\rho^5 }{\left(\rho^2+m_{\psi }^2\right)^{2}}\int _{0}^{\frac{\pi }{2}}d\phi \frac{\sin^{2}\phi\left[\cos^{2}\phi -\left(\sin^{2}\phi\right) /3\right]}{\rho^2 +m_{\phi}^2 +\frac{\rho^{3+2n}\sin^{2+2n}\phi\cos\phi}{M^{1+2n}}},
\end{align}
where in the second line we have used polar coordinates $\omega=\rho \cos\phi, \ k=\rho\sin\phi$ where $\rho$ is the modulus of the (Euclidean) 4-momentum. 
Note that the integration in $\phi$ is now from $0$ to $\pi/2$ ensuring that we are integrating only over the half line with $\omega>0$. 
In the following sections we specify the form of the LI cutoff which is implicit in eq.~\eqref{syin} and which introduces a second scale into the problem.

Before proceeding, we note that the dissipative dispersion relation considered above can actually be seen as an effective LIV regulator for the integral analogous to the ones we considered in section \ref{sec4}; accordingly we could investigate the limit $M\rightarrow\infty$ in the previous equations and study the resulting percolation of LIV without introducing the new LI scale $\Lambda$. In accordance with ref.~\cite{Collins:2004bp} we expect a non-vanishing value of $\Delta c$. In this case, it can be shown that the actual value of $\Delta c$, in this limit, depends on the value of $n$ in the MDR (see appendix~\ref{appF}) similarly to what was observed in section~\ref{sec4} for the dispersive case, and is given by
\begin{equation}\label{asi2} 
\Delta c=-\frac{g^{2}}{48\pi^{2}}\frac{n-1/2}{n+1/2}.
\end{equation}
Note, in addition, that $\Delta c$ vanishes (up to this order) for $n=1/2$.
\subsection{Sharp LI cutoff}\label{sharpdis}

After Wick rotating into Euclidean space, we consider a LI cutoff such as the one discussed in section \ref{sharp}. Accordingly, eq.~\eqref{polar} becomes

\begin{align}\label{disshard}
\Delta c &= -\frac{g^{2}}{ \pi^3} \int _{0}^{\Lambda}d\rho\frac{\rho^5}{\left(\rho^2+m_{\psi }^2\right)^{2}}\int _{0}^{\frac{\pi }{2}}d\phi \frac{\sin^{2}\phi \left[\cos^{2}\phi  -\left(\sin^{2}\phi\right) /3 \right]}{\rho^2 +m_{\phi}^2 +\frac{\rho^{3+2n}\sin^{2+2n}\phi\cos\phi}{M^{1+2n}}}\\ \nonumber
&= -\frac{g^{2}}{\pi^3} \int _{0}^{1}dz\frac{z^5 }{\left(z^2+R_{\psi}\right)^{2}}\int _{0}^{\frac{\pi }{2}}d\phi\frac{\sin^{2}\phi\left[\cos^{2}\phi  -\left(\sin^{2}\phi\right) /3 \right]}{z^2 +R_{\phi} +z^{3+2n}\lambda^{n+1/2}\sin^{2+2n}\phi\cos\phi},
\end{align}
where in the second line we have done the change of variables introduced in eq.~\eqref{hardresc}, i.e. $\rho=\Lambda z$ in order to have dimensionless variables, while we introduced $\lambda=\Lambda^{2}/M^{2}$ and $R_{\phi,\psi}=m^{2}_{\phi,\psi}/\Lambda^{2}$ as we did in section~\ref{sec4}. 

As in section \ref{sec41} the behaviour of $\Delta c$ for $\lambda\ll 1$ can be obtained by expanding the integrand in eq.~\eqref{disshard} around $\lambda=0$. In this case, after plugging this expansion back in eq.~\eqref{disshard} and using that the angular integration of the zeroth-order term vanishes we have
\begin{equation}\label{new2}
\Delta c=-\frac{g^{2}}{\pi^{3}} \frac{2n-1}{3(2n+7)(2n+5)} \lambda^{n+1/2}\int_{0}^{1}\frac{dy}{2}\frac{y^{n+7/2}}{(y+R_{\psi})^{2}(y+R_{\phi})^{2}}+\mathcal{O}(\lambda^{1+2n}),
\end{equation}  
where $y=z^{2}$ (see appendix~\ref{appE} for details on the integral in eq.~\eqref{new2}). Analogously to what we observed after eq.~\eqref{asi2}, $\Delta c$ vanishes (actually at all orders in $\lambda$) for $n=1/2$. 

The percolation $\Delta c\propto (\Lambda/M)^{n+1/2}$ in eq.~\eqref{new2} decreases upon decreasing $\Lambda/M$ analogously to what happens in the pure dispersion case discussed in section \ref{sec4}; the small ratio $\Lambda/M$ controls the behaviour of $\Delta c$ for $\Lambda\ll M$ with the same algebraic power as the one of the LIV scale $M$ in the MDR considered, see eq.~\eqref{prodis}. This demonstrates that, also in the presence of dissipation, the percolation can be tamed by a large separation of scales protecting in this way low-energy physics. Moreover, as in section \ref{sec4}, eq.~\eqref{new2} clearly shows that $\Delta c$ (up to first order in $\lambda^{n}$) is symmetric with respect to the exchange of $R_{\psi}$ and $R_{\phi}$; the limit in which both fields are massless is finite while the dependence of $\Delta c$ on $R_{\psi},R_{\phi}$ is rather weak as long as $m_{\psi},m_{\phi}\ll \Lambda,M$. Finally, we note that the percolation of dissipation is purely real and as such is qualitatively similar to that induced by dispersion.


\subsection{Smooth LI cutoff}
\begin{figure}[tbp]
\centering
\includegraphics[scale=0.6]{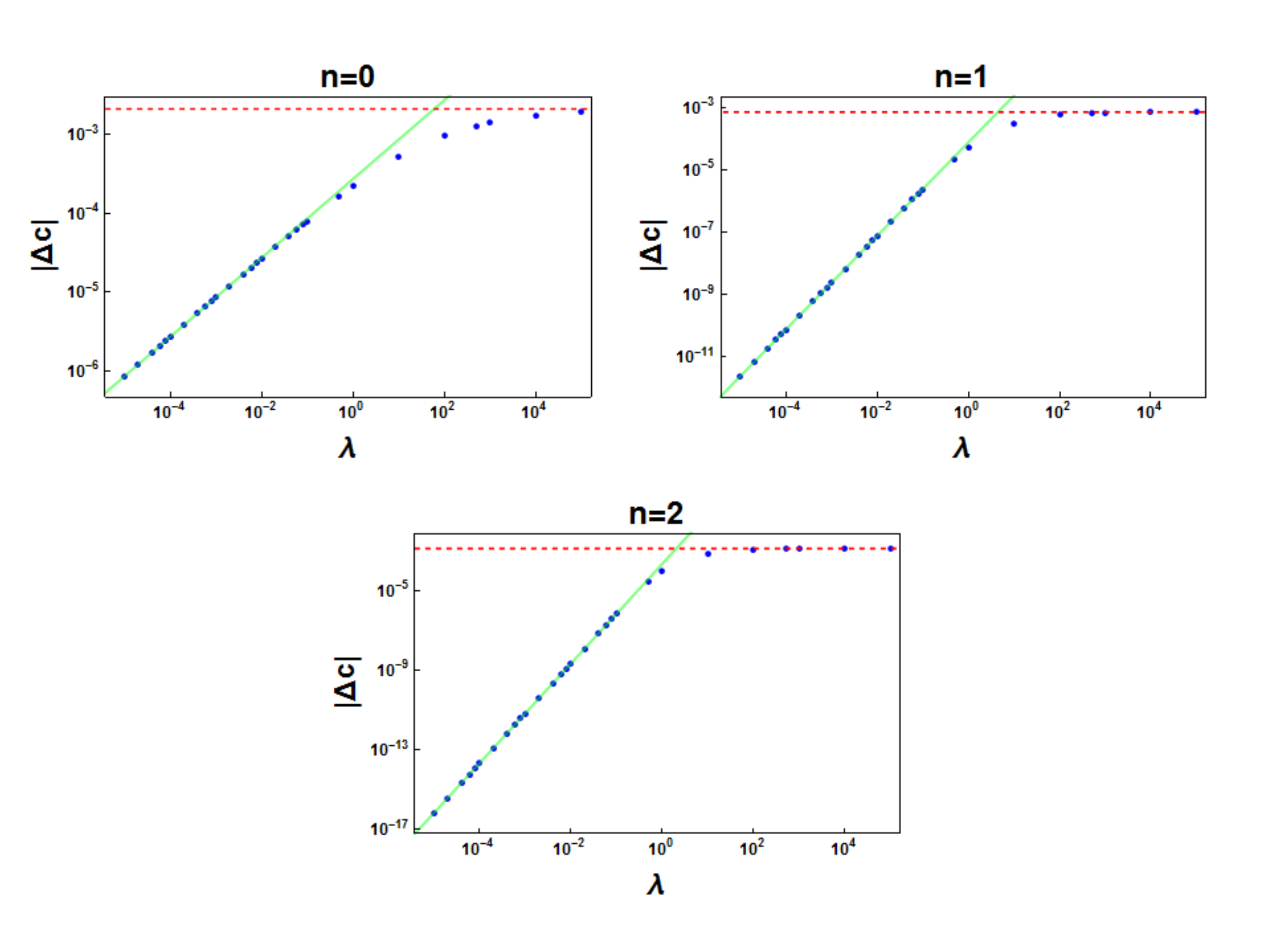}
\caption{Dependence of $|\Delta c|$ on $\lambda\equiv\left(\Lambda/M\right)^{2}$ in the presence of dissipation with a smooth Gaussian cutoff (see eq.~\eqref{dissnl}) with $n=0, 1, 2$ and the same values of the parameters as in figures \ref{hard1} and \ref{nlcut}, i.e. $R_{\phi}=R_{\psi}=10^{-12}$ and $g=1$. Symbols correspond to numerical data while the solid lines correspond to the theoretical prediction $\Delta c\propto (\Lambda/M)^{n+1/2}$ coming from expanding in $\lambda$ for $\lambda\ll 1$. The dashed lines correspond to eq.~\eqref{asi2}, i.e. to the values approached by $\Delta c$ for $\lambda\gg 1$. Note that, these values depend on $n$ while the violation is suppressed for $M\gg\Lambda$..
}\label{dissn0}
\end{figure} 
In this section we consider the case of a smooth LI cutoff, instead of the sharp one investigated in the previous subsection, adopting the same approach as in the dispersive case discussed in section \ref{sec4}. The expression of interest in this case is

\begin{align}\label{dissnl}
\Delta c &= -\frac{g^{2}}{\pi^3} \int _{0}^{\infty}d\rho\; e^{-\rho^{2}/\Lambda^{2}}\frac{\rho^5}{\left(\rho^2+m_{\psi }^2\right)^{2}}\int _{0}^{\frac{\pi }{2}}d\phi\frac{\sin^{2}\phi\left(\cos^{2}\phi -\sin^{2}\phi /3 \right)}{\rho^2 +m_{\phi}^2 +\frac{\rho^{3+2n}\sin^{2+2n}\phi\cos\phi}{M^{1+2n}}}\\ \nonumber
&= -\frac{g^{2}}{\pi^3} \int _{0}^{\infty}dz\ e^{-z^{2}}\frac{z^5 }{\left(z^2+R_{\psi}\right)^{2}}\int _{0}^{\frac{\pi }{2}}d\phi\frac{\sin^{2}\phi\left(\cos^{2}\phi  -\sin^{2}\phi /3 \right)}{z^2 +R_{\phi} +z^{3+2n}\lambda^{\frac{1+2n}{2}}\sin^{2+2n}\phi\cos\phi},
\end{align}
where the smooth cutoff is introduced in the Euclidean space via the Gaussian function with parameter $\Lambda$, see also eq.~\eqref{nonlocal}. On the second line we introduced the same parameters as in eq.~\eqref{nonlocal}. Repeating the analysis done before we conclude that also in this case $\Delta c\propto (\Lambda/M)^{n+1/2}$ and therefore $\Delta c$ is suppressed with a wide separation of scales $\Lambda\ll M$.
The only difference compared to eq.~\eqref{new2} is in the radial integration in $y$ which runs up to $\infty$ after the introduction of the cutoff function, as in eq.~\eqref{change}. We have computed numerically the integral eq.~\eqref{dissnl} as a function of $\lambda=\Lambda/M$ for various values of $n$. The result for $\Delta c$ with $R_{\phi,\psi}=10^{-12}$ is shown in figure \ref{dissn0} together with its asymptotic behaviours.
As in the dispersive case, we have confirmed that the value approached by $\Delta c$ for large $\lambda$, i.e. for $\Lambda\gg M$ with fixed $R_{\phi,\psi}$ is given by eq.~\eqref{asi2} both for sharp and smooth cutoffs, in agreement with the values approached when only the MDR is present as a regulator and with the analytical argument presented in appendix~\ref{appF}. This agrees  with the result of ref.~\cite{Collins:2004bp} of unsuppressed LIV percolation. 

\section{Summary and discussion}\label{sec6}

In this work we have revisited the problem of the IR percolation of Lorentz invariance violations due to radiative corrections. We have considered a model consisting of a scalar and a fermionic field coupled by a Yukawa interaction, which was already used in the literature for studying the naturalness problem of LIVs and which provides a cartoon of the structure of the scalar sector of the standard model of particle physics.

In sections~\ref{sec2} and~\ref{section3}, we discussed the instance in which only the UV Lorentz breaking scale (usually identified with the Planck scale in the quantum gravity literature) is present in the problem in addition to the mass scales of the particles. In this case one can model the LIV by introducing a Lorentz-breaking cutoff which eliminates large momenta in a certain preferred reference frame. This is the way the problem was tackled in previous studies~\cite{Urrutia:2005ty}. We revisited the original calculation of ref.~\cite{Collins:2004bp}, making explicit various steps as well as considering the percolation of LIV on the fermion field. In this case, our results agree with ref.~\cite{Urrutia:2005ty}, though our analysis is more general. Indeed, we make no assumption concerning the masses of the fields and we consider also a cutoff for both the fermionic and scalar fields. We emphasize the interesting results reported in section~\ref{32}, which show a possible cancellation of the LIV percolation even in rather generic situations. Indeed, we have shown that if the cutoff function is the same for both these fields, the percolation on the fermion is completely suppressed, at least up to one loop in perturbation theory. Clearly, unsuppressed percolation are anyhow present on the scalar due to fermionic loops. In this respect, the case with unequal scalar and fermion masses $m_{\phi}$ and $m_{\psi}$, respectively, and in particular with $m_{\phi}\gg m_{\psi}$ is the most physically interesting. Indeed, in an EFT approach, the heavy scalar has to be traced out for describing low-energy physics which, in this way, becomes unaffected by the unsuppressed percolation on the scalar field, at least at this order in perturbation theory.

In sections~\ref{sec4} and~\ref{sec5} we have considered the case in which a scale $M$ of LIV and a scale $\Lambda$ are present, where $\Lambda$ is a LI cutoff representing the scale of validity of the EFT. In particular, we assume that between such a  Planck scale $M$ (setting the Lorentz breaking) and the low-energy physics there exist some extra (per se, see section~\ref{introduction}) Lorentz invariant physics. The LI cutoff $\Lambda$ has been introduced via both a sharp and a smooth cutoff function in Euclidean momentum space. We have shown that if these two scales are well separated, i.e. $\Lambda\ll M$, the percolation is suppressed. While this result could have been expected on mathematical and physical ground, we determine here the scaling behaviour of the percolation for various MDR: In particular in section \ref{sec4} we consider the case in which $\Lambda/M\gg 1$ with various modified dispersion relations $\Delta$ given by eq.~\eqref{MDR}. Heuristically this case is expected to be equivalent to the one investigated in refs.~\cite{Collins:2004bp,Urrutia:2005ty} due to the fact that the MDRs themselves act as cutoffs. Indeed, it can be seen from eq.~\eqref{asi} that we find an unsuppressed percolation but the value of the corresponding $\Delta c$ (up to one loop) depends on the detail of the MDR, a fact that was not noticed in previous discussions. Secondly, we consider the case of $\Lambda/M\ll 1$ and show that the percolation depends linearly on $\Lambda/M$ to the power with which $M$ appears in the MDR, see figures~\ref{hard1} and~\ref{nlcut}, a fact which is quite interesting from the phenomenological point of view. Indeed, most available models of quantum gravity seem to preserve CPT symmetry: this means that CPT odd operators appearing in the MDR are disfavored whereas CPT even mass dimension 5 or 6 operators can be shown to give negligible contributions or terms of the type $p^{4}/M^{2}$ \cite{Liberati:2013xla}. In this case, we have shown that one could expect a possibly strong suppression of the form $\Lambda^{2}/M^{2}$. If finally $M$ is identified with the Planck mass and we consider a constraint on the LIV dimensionless parameter of the order of $10^{-18}$ (from constraints on the neutrino-electron sector of the SM, see ref.~\cite{Borriello:2013ala}) we see that the EFT scale $\Lambda$ has to be simply less than $10^{10}$ GeV for evading current constraints.

Last but not least, we have considered also the instance in which the scalar field is affected by dissipation. This case has not been treated extensively in the literature in spite of the fact that interesting phenomenology can be extracted by allowing dissipative MDR (see refs.~\cite{Parentani:2007uq,Liberati:2013usa}) and that from the theoretical point of view dissipation must be present if LIV arises dynamically \cite{Parentani:2007uq}. In section~\ref{sec5} (see eq.~\eqref{polar} and the discussion thereof) we found the unexpected result that no percolation of these dissipative effects occurs, i.e. that no imaginary contribution emerges as a consequence of the LIV percolation, neither in $\Delta c$ of the fermion nor in its mass, as the one-loop correction to it turns out to be real (see appendix~\ref{appD}). Accordingly, the mass correction does not imply a stringent constraint on the LIV percolation. Moreover, we also show that a separation of scales suppresses again the LIV percolation in the same fashion as it does in the purely dispersive case of section \ref{sec4}, see figure \ref{dissn0}.
While the idea that separation of scales can prevent unsuppressed LIV is not entirely new, our work highlights in a pedagogical way how such a mechanism works while extending it to the interesting case of field theories with effective dissipation. We hope that the present work will stimulate further investigation in this direction in a near future.


\acknowledgments

The authors would like to thank David Mattingly for comments and discussions. SL wants to thank Luca Maccione for stimulating discussions that led to this project. AB would like to thank Marco Letizia, Mauro Valli and Bruno Lima de Souza for interesting discussions. This publication was made possible through the support of the grant No.~51876 from the John Templeton 
Foundation. The opinions expressed in this publication are those of the authors and do not 
necessarily reflect the views of the John Templeton Foundation.

\bibliographystyle{JHEP}
\bibliography{mylibrary}

\appendix

\section{Calculation of traces
\label{appA}}
Here we calculate the trace appearing on the last line of eq.~\eqref{cab} which determines the values of $c_{aa}$. In order to do so we use the following relations involving the LIV fermion propagator $G$, the standard fermion propagator $S_{0}$, the cutoff function $f(\textbf{k}^{2}/\Lambda^2)$ and the LIV cutoff $\Lambda$ (these quantities are introduced in the main text after eq.~\eqref{popo}): 
\be
\frac{\partial G(k)}{\partial k^{a}}=\frac{\partial S_{0}(k)}{\partial k^{a}} f+ S_{0}(k)\frac{\partial f(\textbf{k}^{2}/\Lambda^2)}{\partial k^{a}},
\ee

\be
\frac{\partial S_{0}(k)}{\partial k^{a}}=i \eta_{aa}\left(\frac{\gamma^{a}}{k^{2}-m^{2}}-2k^{a}\frac{\slashed{k}+m}{(k^{2}-m^{2})^{2}}\right),
\ee

\be
\frac{\partial f}{\partial k^{a}}=\frac{2k^{a}}{\Lambda^{2}}f'(\textbf{k}^{2}/\Lambda^{2})(1-\delta_{a0}),
\ee
where $\eta_{ab}$ is the Minkowski metric and $\delta_{ab}$ is the usual Kronecker delta.
For simplicity, in these expressions we do not write explicitly the $+i\epsilon$ factors which characterises the denominators of propagators, as they are inconsequential for the present discussion.   
Using the above relations it is possible to express explicitly $\partial G(k)/\partial k^{a}$. Inserting this expression in the trace and using the properties of gamma matrices we end up with
\begin{align}\label{trace}
tr\left[\frac{\partial G(k)}{\partial k^{a}}\frac{\partial G(k)}{\partial k^{a}}\right]&=16 (k^{a})^{2}(k^{2}+m^{2})\\ \nonumber
& \cdot\left[\frac{f'\,(1-\delta_{a0})}{\Lambda^2(k^{2}-m^{2})^{2}}+\frac{f^{2}}{(k^{2}-m^{2})^{2}}-\frac{2f'\,f}{\Lambda^{2}(k^{2}-m^{2})^{3}}\eta_{aa}(1-\delta_{a0})\right]\\ \nonumber
&+4\eta_{aa}\frac{f^{2}}{(k^{2}-m^{2})^{2}}\\ \nonumber
&+16(k^{a})^{2}\frac{f\,\eta_{aa}}{(k^{2}-m^{2})}\left[\frac{f'(1-\delta_{a0})}{(k^{2}-m^{2})}-\eta_{aa}\frac{f}{(k^{2}-m^{2})^{2}}\right],
\end{align}
which is used in section \ref{dd}.

\section{Useful integrals for sections \ref{sec2} and \ref{section3}} 
\label{appB}
Here we collect some formulas which are used throughout the paper, in particular in eqs.~\eqref{cab} and~\eqref{gen} in order to derive eqs.~\eqref{c00}, \eqref{c11}, \eqref{p} and \eqref{q}.

In particular, concerning the integrals in section \ref{sec2},  let us indicate
\begin{equation}
I_{n}\equiv\int_{-\infty}^{\infty}\frac{dk^{0}}{2\pi}\frac{1}{((k^{0})^{2}+\textbf{k}^{2}+m^{2})^{n}}=\int_{-\infty}^{\infty}\frac{dk^{0}}{2\pi}\frac{1}{((k^{0})^{2}+A)^{n}},
\end{equation}
where we are working in the Euclidean space, where $A\equiv \textbf{k}^{2}+m^2$ and $\textbf{k}$ is the spatial three momentum. These kind of integrals are needed for calculating $c_{11}$ according to eq.~\eqref{cab}, taking into account also eq.~\eqref{trace}. They can be explicitly calculated and in particular one finds
\begin{align}\label{intc1}
I_{1}=\frac{1}{2\sqrt{A}},\quad I_{2}=\frac{1}{4A^{3/2}}, \quad  I_{3}=\frac{3}{16A^{5/2}}, \quad I_{4}=\frac{5}{32A^{7/2}}. 
\end{align}
For calculating  $c_{00}$ from the same expressions, instead, we need also the following integrals which, in turn, can be expressed in terms of the previous ones:
\begin{align}\label{intc0}
&\int\frac{dk^{0}}{2\pi}\frac{(k^{0})^{2}}{((k^{0})^{2}+A)^{3}}=I_{2}- A\,I_{3}=\frac{1}{16A^{3/2}},\\ \nonumber
&\int\frac{dk^{0}}{2\pi}(k^{0})^{2}\frac{k^{2}-m^{2}}{((k^{0})^{2}+A)^{3}}=(I_{2}- A\,I_{3})-2m^{2}(I_{3}-A\,I_{4})=\frac{1}{16A^{3/2}}(1-m^{2}A^{-1}),
\end{align}
which are used in deriving eq.~\eqref{c00}.

Concerning section \ref{section3}, instead, in order to calculate the $k^{0}$ integration which leads to eqs.~\eqref{p} and \eqref{q} from eq.~\eqref{gen} we have used the following identities,
\begin{align}
& \int_{-\infty}^{\infty} \frac{dk^{0}}{2\pi} \frac{(k^{0})^{2}}{((k^{0})^{2}+A)^{2}((k^{0})^{2}+B)}=\frac{1}{4\sqrt{A}(\sqrt{A}+\sqrt{B})^{2}},\\
& \int_{-\infty}^{\infty} \frac{dk^{0}}{2\pi} \frac{1}{((k^{0})^{2}+A)^{2}((k^{0})^{2}+B)}=\frac{(2\sqrt{A}+\sqrt{B})}{4A^{3/2}(\sqrt{A}+\sqrt{B})^{2}\sqrt{B}},\\
& \int_{-\infty}^{\infty} \frac{dk^{0}}{2\pi} \frac{1}{((k^{0})^{2}+A)^{2}((k^{0})^{2}+B)}=\frac{1}{2 \left(A \sqrt{B}+\sqrt{A} B\right)}, 
\end{align} 
where $A=\textbf{k}^{2}+m_{\psi}^{2}$ and $B=\textbf{k}^{2}+m_{\phi}^{2}$. Note that these relations hold also for the case with modified dispersion relations (MDR) discussed in section \ref{sec4} with suitable corresponding definition of $A$ and $B$.


 
\section{Asymptotic behaviours}
\label{appF}
In this appendix we demonstrate eqs.~\eqref{asi} and~\eqref{asi2} concerning the asymptotic behaviour of $\Delta c$ for $\ m_{\psi,\phi}\ll M$ in the case in which the scale $M$ of the MDR (see eq.~\eqref{MDR}) is the only cutoff scale present, with the additional assumption that $\Lambda\gg M$ when the LI cutoff scale $\Lambda$ is also present.

\subsection{Dispersive case} 
Let us consider  eq.~\eqref{hardsph}, rescale the momenta by $M$ and consider the case in which $\Lambda\gg M$ and $\ m_{\psi,\phi}\ll M$; for convenience hereafter we assume $m_{\psi}=m_{\phi}=0$, then we have
\begin{equation}\label{eI}
\Delta c=-\frac{g^{2}}{4\pi^3}\int_{0}^{\infty}dy\int_{0}^{\pi}d\phi\;\sin^{2}\phi\;\frac{\cos^{2}\phi-(\sin^{2}\phi)/3}{y+y^{1+n}\sin^{2+2n}\phi}\equiv -\frac{g^{2}}{4\pi^3}I,
\end{equation}
where we defined $y=k^{2}/M^{2}$. The integral on $y$ runs up to $\infty$ both in the case with a sharp and smooth cutoff as we are anyhow interested in the limit $\Lambda/M\gg 1$. Moreover we will see that the result is actually determined by the behaviour of the integrand  around $y=0$ which is not affected by the way the cutoff is imposed. The integral $I$ defined in eq.~\eqref{eI} can be regularised by introducing a $\epsilon >0$
\begin{equation}\label{I}
I=\lim_{\epsilon\rightarrow 0}\int_{\epsilon}^{\infty}dy\int_{0}^{\pi}d\phi \frac{\mathcal{P}(\phi)}{y+y^{1+n}C},
\end{equation}
where we defined $C=\sin^{2+2n}\phi$ and 
\begin{equation}\label{Pf}
\mathcal{P}(\phi)=\sin^{2}\phi\left[\cos^{2}\phi-(\sin^{2}\phi)/3\right].
\end{equation} 
Note that $C>0$ and 
\begin{equation}\label{po}
\int_{0}^{\pi}d\phi\,\mathcal{P}(\phi) = 2 \int_0^{\pi/2} d\phi\,\mathcal{P}(\phi) =0.
\end{equation} 
Let us first consider the integral over $y$, which is logarithmically divergent as $\epsilon\rightarrow 0$. In fact 
\begin{align}\label{decompo}
\int_{\epsilon}^{\infty}dy\frac{1}{y+y^{1+n}C}&=\int_{\epsilon C^{1/n}}^{\infty}d\chi\frac{1}{\chi+\chi^{1+n}}\\ \nonumber
&= -\log\epsilon + \int_{0}^{1}d\chi\left(\frac{1}{\chi+\chi^{1+n}}-\frac{1}{\chi}\right)+\int_{1}^{\infty}d\chi\frac{1}{\chi+\chi^{1+n}} -\frac{1}{n}\log C + {\mathcal O}(\epsilon),
\end{align} 
where on the first line $\chi=C^{1/n}y$, while on the second one we have taken the limit $\epsilon\rightarrow 0$ of the first integral since it is finite for $n>0$. The first three terms on the last line do not depend on $\phi$ and due to eq.~\eqref{po}, they do not contribute to $I$ in eq.~\eqref{I}, while the last term yields 
\begin{equation}\label{Anew}
I=-2\frac{n+1}{n}\int_{0}^{\pi}d\phi\,  \mathcal{P}(\phi) \log(\sin\phi).
\end{equation}
This integral can be calculated by using its symmetry around $\phi=\pi/2$, the change of variable $x=\sin\phi$, and the fact that $\log x=\lim_{\epsilon\rightarrow 0}(x^{\epsilon}-1)/\epsilon$, which allows us to express $I$ in terms of
\begin{equation}
i_\alpha \equiv \int_0^1 dx\frac{x^\alpha}{\sqrt{1-x^2}} = \frac{\sqrt{\pi}}{2}\frac{\Gamma\left(\frac{1+\alpha}{2}\right)}{\Gamma\left(1+\frac{\alpha}{2}\right)} \qquad (\alpha > -1),
\end{equation}
and conclude that
\begin{equation}
I = \frac{n+1}{n} \frac{\pi}{12}.
\end{equation}
Accordingly,
\begin{equation}
\Delta c=-\frac{g^{2}}{48\pi^2}\frac{n+1}{n},
\end{equation} 
in accordance with eq.~\eqref{asi}. 
Note that, as anticipated, the integral $I$ is essentially determined by the algebraic singularity of the integrand as a function of $y\to 0$ and this shows that the eventual result is independent of how the LI cutoff $\Lambda$ is actually imposed, as this affects the integrand only at $y\simeq 1$.

\subsection{Dissipative case}
Equation~\eqref{asi2} can be derived from eq.~\eqref{polar} by following the same steps as above. Indeed, eq.~\eqref{polar} can be written as (as we are interested in the IR limit we set $m_{\psi,\phi}=0$) 
\begin{equation}
\Delta c=-\frac{g^{2}}{2\pi^3}\int_{0}^{\infty}dy\int_{0}^{\pi/2}d\phi  \frac{\mathcal{P}(\phi)}{y+y^{3/2+n}\sin^{2+2n}\phi\cos\phi}\equiv -\frac{g^{2}}{2\pi^3}\tilde{I},
\end{equation}
where we defined $y=k^{2}/M^{2}$ and $\mathcal{P}(\phi)$ is given in eq.~\eqref{Pf}. Defining now $\sin^{2+2n}\phi\cos\phi\equiv C$ we see that the $y$-integration in $\tilde{I}$ is formally analogous to the one in $I$ defined in eq.~\eqref{I} and has a decomposition similar to eq.~\eqref{decompo} with $n\rightarrow n+1/2$. Then, using eq.~\eqref{po} we remain with only the angular part to be computed. This is given by
\begin{equation}
\int_{0}^{\pi/2}d\phi \mathcal{P}(\phi) \left[(2n+2)\log(\sin\phi)+\log(\cos\phi)\right].
\end{equation}
The first term is the one already computed in eq.~\eqref{Anew}, giving $-(2n+2)\pi/48$ while the second term is analogously evaluated to be $\pi/16$. The final result is then 
$$
\Delta c=-\frac{g^{2}}{2\pi^{2}(n+1/2)}\left(\frac{2n-1}{48}\right)=-\frac{g^{2}}{48\pi^{2}}\frac{n-1/2}{n+1/2},
$$
in accordance with eq.~\eqref{asi2}. Again, as before, the entire contribution to the integral $\tilde{I}$ comes from the region of integration around $y=0$ and this shows that the result is actually independent of how the LI cutoff $\Lambda$ is imposed.
\section{Useful integrals for sections \ref{sec4} and \ref{sec5}} 
\label{appE}

In this appendix we report the results for the integrals in eqs.~\eqref{new1} and~\eqref{new2}.
In particular for the former we need to determine the following integral
\begin{equation}\label{dispan}
Q_n(R_\phi,R_\psi) \equiv  \int_{0}^{1}\frac{dy}{2}\frac{y^{3+n}}{(y+R_{\psi})^{2}(y+R_{\phi})^{2}},
\end{equation}
where $R_{\psi,\phi}\geq 0$. 
Note that $Q_n$ can be calculated as
\begin{equation}
\label{rdiv}
Q_n(R_\psi,R_\phi) = \frac{\partial^2}{\partial R_{\psi}\partial R_{\phi}} \frac{S_n(R_\psi) - S_n(R_\phi)}{R_{\phi}-R_{\psi}} \quad \mbox{where} \quad 
S_n(R_{\phi,\psi}) \equiv \int_{0}^{1}\frac{dy}{2} \frac{y^{3+n}}{y+R_{\phi,\psi}};
\end{equation}
the remaining integral $S_n$ is simpler and can be expressed in terms of hypergeometric $_{2}F_{1}$ functions. However, for convenience, we omit the lengthy explicit expression of the resulting $Q_n(R_\psi,R_\phi)$, which can straightforwardly determined as explained above. Here we only note that when both fields are massless, i.e. $R_{\psi,\phi}=0$ its expression is particularly simple: $Q_n(0,0) =1/(2n)$.

Concerning eq.~\eqref{new2}, instead, we would like to calculate the following integral
\begin{equation}\label{dissan}
\int_{0}^{1}\frac{dy}{2}\frac{y^{n+7/2}}{(y+R_{\psi})^{2}(y+R_{\phi})^{2}},
\end{equation}
which actually corresponds to $Q_{n+1/2}(R_\phi,R_\psi)$ defined above. Accordingly, it can also be expressed in terms of $S_{n+1/2}(R_{\phi,\psi})$ as in eq.~\eqref{rdiv} and, in turn, in terms of hypergeometric $_{2}F_{1}$ functions. For convenience, we do not report its lengthy expression here but only quote the value eq.~\eqref{dissan} takes in the massless case $R_{\phi,\psi}=0$, i.e. $Q_{n+1/2}(0,0) =1/(2n +1)$.

\section{Reality and pole structure in the dissipative case
\label{appD}}
In this appendix we analyze the location of the poles and the possibility of performing the Wick rotation in section \ref{sec5}. Moreover, we discuss also the correction to the mass of the fermion arising in the dissipative case due to its self-energy.

First, we study the location of the poles of the integrand and the reality of the following integral
\begin{equation}\label{d1}
i \int_{0}^{\infty} d\omega \frac{\omega^{2}+\frac{k^{2}}{3}}{(\omega^2 - k^{2}-m_{\psi}^2+i\epsilon)^{2}\left(\omega^2 - k^{2}-m_{\psi}^2+i\epsilon+i\omega\frac{k^{2+2n}}{M^{1+2n}}\right)},
\end{equation}
which is relevant for the discussion in section \ref{sec5} (see eq.~\eqref{syin}).
Note that the primitive of this integral (which does not show any singularity within the domain of integration, as long as $\epsilon\neq 0$) can be calculated explicitly (however, we do not report here its lengthy expression) and one can show that in the limit $\epsilon\rightarrow 0^{+}$ the corresponding integral is indeed finite and real. In addition its subsequent integration in $k$ (see eq.~\eqref{syin}) is finite and therefore this limit can be taken from the outset. The location of the poles of the integrand in the complex $\omega$-plane, is easy to determine by studying the zeros of the denominator. Doing so it turns out that the poles are always located outside the region with both ${\rm Re}[\omega]$ and ${\rm Im}[\omega]$ positive in such a way that the integral on a close contour entirely within this region gives zero due to Cauchy's theorem. Now given the structure of the integrand it is easy to see that on the arch of circumference of large radius $\Omega$ which lies within that first quadrant, the integral is bounded by  $\approx 1/\Omega^3$, so that it vanishes in the limit $\Omega\rightarrow\infty$. Then we can Wick rotate the integral in eq.~\eqref{d1}, obtaining finally    
\begin{equation}
-\int_{0}^{\infty} d\omega \frac{\omega^{2}-\frac{k^{2}}{3}}{(+\omega^2 + k^{2}+m_{\psi}^2)^{2}\left(\omega^2 + k^{2}+m_{\psi}^2+\omega\frac{k^{2+2n}}{M^{1+2n}}\right)},
\end{equation}
which is used for deriving eq.~\eqref{polar}.

\subsection*{Mass correction}
The same reasoning as above can be applied for studying the correction to the fermion mass. The latter is encoded in $\chi_{1}(0)$ of eq.~\eqref{ol} (considering in this case a dissipative term in the scalar propagator, see eq.~\eqref{prodis}, with $f=\tilde{f}=1$), i.e.
\begin{equation}\label{ma}
-ig^2 \int_{0}^{\infty}\frac{dk}{(2\pi)^3}4\pi k^{2}\int_{-\infty}^{\infty}\frac{d\omega}{2\pi}\frac{1}{(\omega^2-k^2-m_{\psi}^{2}+i\epsilon)(\omega^{2}-k^2 -m_{\phi}^2+i\epsilon+i|\omega|k^{2+2n}/M^{1+2n})}.
\end{equation}
The symmetry under $\omega\rightarrow -\omega$ of the integrand and the same splitting of the integral as the one invoked in deriving eq.~\eqref{syin} can be used in this expression in order to arrive at
\begin{equation}\label{ma2}
-8\pi i g^2 \int_{0}^{\infty}\frac{dk}{(2\pi)^3}k^{2}\int_{0}^{\infty}\frac{d\omega}{2\pi}\frac{1}{(\omega^2-k^2-m_{\psi}^{2}+i\epsilon)(\omega^{2}-k^2 -m_{\phi}^2+i\epsilon+i\omega k^{2+2n}/M^{1+2n})}.
\end{equation}
At this point it is again possible to calculate the primitive of the $\omega$-integral and show that in the limit $\epsilon\rightarrow 0$ the whole expression is real. The poles of the integrand are located at the same positions as above and this permit again to perform the Wick rotation. The successive integration in $k$ is finite (apart from the case in which both fields are massless for which there is an IR divergence, as expected). This shows that the mass does not acquire an imaginary part (at least at one loop). 

\end{document}